\documentclass[final,fleqn]{cas-dc}
\usepackage{geometry}
\usepackage[authoryear]{natbib}
\usepackage{subcaption}
\usepackage{tabularx}
\usepackage{dashrule}
\usepackage{tikz}
\usepackage{pgfplots}
\pgfplotsset{compat=1.18}
\usetikzlibrary{patterns}
\usepackage{placeins}
\usepackage{adjustbox}

\def\tsc#1{\csdef{#1}{\textsc{\lowercase{#1}}\xspace}}
\tsc{AVSR}
\tsc{WER}
\tsc{AV}
\begin{document}

\shorttitle{Improving AVSR through Synthetic Visual Data Augmentation}

\shortauthors{Buitrago et~al.}

\title[mode=title]{Improving Audiovisual Speech Recognition through Synthetic Visual Data Augmentation}

% [PLACEHOLDER: add \tnotemark[...] here if a title footnote is needed]

% Title footnote text
% [PLACEHOLDER: add \tnotetext[...]{...} here if acknowledgements/funding
%  notes about the title are needed, e.g. grant numbers, conference version, etc.]

\author[1,2]{Pol Buitrago}[orcid=0009-0005-1055-129X]
\cormark[1]
\ead{pol.buitrago@bsc.es}
\credit{Conceptualization, Methodology, Software, Investigation, Formal analysis, Writing -- original draft}
\author[1]{Pol Gálvez}
\credit{Data curation, Writing -- review \& editing}
\author[1]{Javier Hernando}
\credit{Conceptualization, Supervision, Writing -- review \& editing}

\affiliation[1]{organization={Universitat Politècnica de Catalunya (UPC)},
    addressline={Carrer de Jordi Girona, 1--3},
    city={Barcelona},
    postcode={08034},
    country={Spain}}

\affiliation[2]{organization={Barcelona Supercomputing Center (BSC)},
    addressline={Carrer de Jordi Girona, 29},
    city={Barcelona},
    postcode={08034},
    country={Spain}}
    
\cortext[cor1]{Corresponding author}

% [PLACEHOLDER: \fntext[fn1]{...} for any author footnotes]

% [PLACEHOLDER: \nonumnote{...} if an unnumbered note is needed]

\begin{abstract}
Audiovisual Speech Recognition (AVSR) is a multimodal approach to speech recognition that incorporates visual information from lip movements to enhance model performance. Despite its advantages, its development remains constrained by the limited availability of labeled audiovisual (AV) datasets. This work explores the use of synthetic visual data as a solution, using an audio-driven talking-head pipeline to generate lip-synchronized visual content from existing audio data. We evaluate the effectiveness of synthetic visual data both as an augmentation strategy and as a standalone training resource, applying our approach to Spanish and Catalan. Our results show that augmenting real AV data with synthetic samples yields relative Word Error Rate (WER) reductions of up to 16.2\%, demonstrating the potential of this approach. Moreover, we demonstrate that synthetic data alone can serve as a baseline for AVSR training in languages lacking AV datasets. These findings provide evidence that synthetic visual data can serve as a scalable solution to AVSR data scarcity, enabling broader language coverage.
\end{abstract}

% Keywords
% Each keyword is separated by \sep

\maketitle

\section{Introduction}\label{sec:introduction}
    Audiovisual Speech Recognition (AVSR) is a specialized field within speech-to-text technology that improves transcription accuracy by incorporating visual speech cues alongside the acoustic signal. Unlike unimodal automatic speech recognition (ASR) systems, which rely exclusively on the audio signal, AVSR leverages a multimodal strategy that integrates lip movements and facial cues to enhance performance, particularly in noisy environments where audio data alone is insufficient \citep{Sumby1954VisualCT, xia2020avsrreview, shi2022avhubert, shi2022avsr, zhu2024avwav2vec2}.
    
    Although AVSR offers significant benefits over unimodal recognition approaches, its widespread development across languages is constrained by the limited availability of speech-annotated audiovisual (AV) datasets. While large-scale audio-only corpora are widely available, paired audiovisual speech data remains scarce, especially for under-resourced languages \citep{czyzewski2017audiovisual,brigato2020smalldata, bansal2022datascarcity, dai2022ciavsr,anwar2023muavic}.
    
    State-of-the-art AVSR systems commonly rely on self-supervised pre-training over large-scale unlabeled audiovisual data, followed by supervised fine-tuning on labeled AV corpora \citep{pan2022leveraging, shi2022avhubert, shi2022avsr, anwar2023muavic, zhu2024avwav2vec2}. While this paradigm has substantially reduced annotation requirements \citep{devlin2019bert}, it does not eliminate the need for labeled AV data during fine-tuning, which remain scarce in many languages. This data bottleneck limits the development of robust AVSR systems across languages, underscoring the need for approaches that reduce dependence on real audiovisual annotations.
    
    This work proposes leveraging synthetic visual data paired with real audio to address the scarcity of resources for languages with limited AV datasets. To this end, we introduce a novel speech-driven visual synthesis pipeline for generating synthetic audiovisual training data. By generating visual speech sequences from existing speech recordings and face images \citep{goodfellow2014gan, Prajwal2020Wav2Lip}, we aim to construct scalable AV training data for AVSR systems. This approach reduces the dependence on manually annotated video corpora while enabling the expansion of training resources to under-resourced languages.
    
    Recent advances in generative modeling have made this approach feasible by enabling the synthesis of realistic and temporally coherent visual speech from audio signals \citep{Prajwal2020Wav2Lip,vougioukas2020realistic,li2022speechdriven,rakesh2025talkingheadreview}. These models have demonstrated the ability to generate lip movements that preserve meaningful articulatory structure \citep{shan2022speechinnoise,varano2022speechdriven,agarwal2023moocs,yu2024syntheticfaces}, supporting their use in visual speech tasks such as lipreading \citep{varano2022speechdriven}.
    
    Synthetic visual data has predominantly been explored in the context of unimodal lipreading, where it has shown potential to improve visual speech recognition performance \citep{liu2023synthvsr, yang2024audiovsr}. However, whether these benefits transfer to audiovisual speech recognition remains an open question. Unlike visual-only systems, AVSR requires learning coherent joint representations from synchronized audio and visual streams. Consequently, synthetic visual data that is sufficiently informative to improve lipreading performance may still be unsuitable for AVSR if it introduces domain shift, or audiovisual misalignment, which can interfere with cross-modal learning, and degrade overall AVSR performance.
        
    The aim of this work is to investigate the effectiveness of synthetic visual data for AVSR under two complementary scenarios. First, we evaluate its use as a data augmentation strategy in a low-resource audiovisual setting (Spanish). Second, we assess whether it can serve as a standalone training resource in the absence of real audiovisual data (Catalan).

\section{Methodology}\label{sec:methodology}
    \begin{figure*}[t!]
        \centering
        \includegraphics[width=\linewidth]{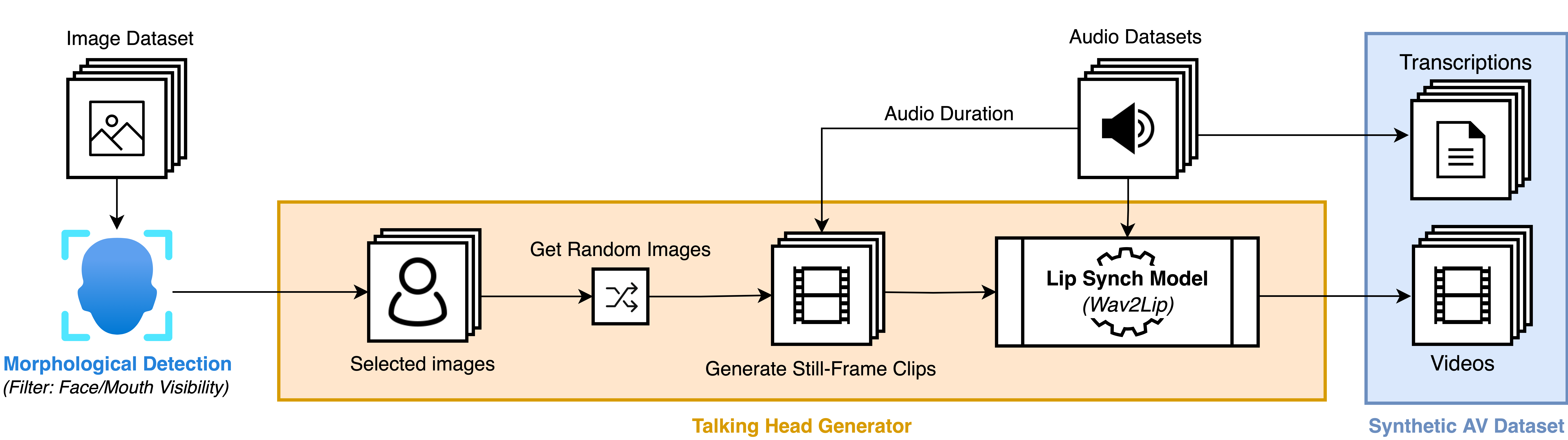}
        \caption{Audio-Visual Synthetic Data Generator (\textit{AVSynthGen}) pipeline scheme. The system generates synchronized audiovisual data by applying lip synchronization to static images using speech audio inputs.}
        \label{fig:speech_production}
    \end{figure*}
    
    To explore the effectiveness of synthetic visual data for AVSR, our methodology is organized into three main components: (i) a speech-driven visual synthesis pipeline that generates lip-synced AV content from face images and speech recordings, (ii) fine-tuning a pretrained AVSR model on real, synthetic, and mixed datasets, and (iii) a robust experimental setup to evaluate the impact of synthetic data on model performance.

\subsection{Speech-Driven Visual Synthesis}\label{subsec:pipeline}
    To address the issue of audiovisual data scarcity, we propose a novel speech-driven visual synthesis pipeline, named \textbf{\textit{AVSynthGen}}, an automated system that generates lip-synchronized audiovisual data by combining speech audio with face images and animating the mouth region. The pipeline produces synthetic visual sequences temporally aligned with real speech, resulting in artificial audiovisual samples tailored for lipreading and audiovisual speech recognition tasks. An overview of the generation process is shown in Figure~\ref{fig:speech_production}.

    \begin{figure}[pos=b]
        \centering
        \includegraphics[
            width=0.9\columnwidth,
            trim={0 -20 0 0},
            clip
        ]{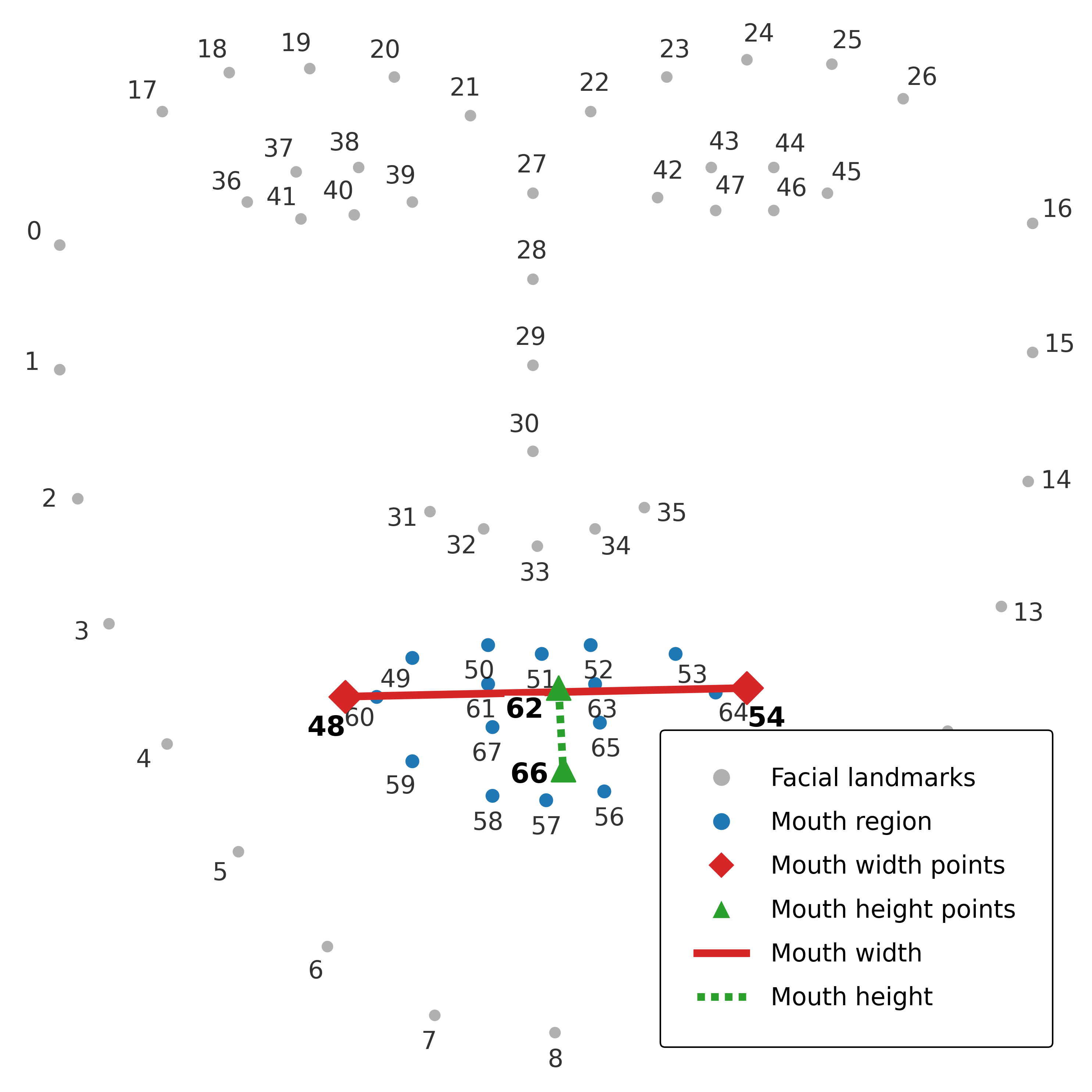}
        \caption{Visualization of the 68 facial landmark template used in the morphological filtering stage. The facial landmarks are extracted using dlib \citep{king2009dlib}.}
        \label{fig:landmark_template}
    \end{figure}
    
\begin{figure}[pos=t]
    \centering
    \begin{adjustbox}{max width=\columnwidth}
        \begin{minipage}{\linewidth}
            \centering
            \begin{subfigure}[t]{0.415\columnwidth}
                \centering
                \makebox[\linewidth][c]{%
                    \hspace*{0.5em}%
                    \includegraphics[width=0.97\linewidth]{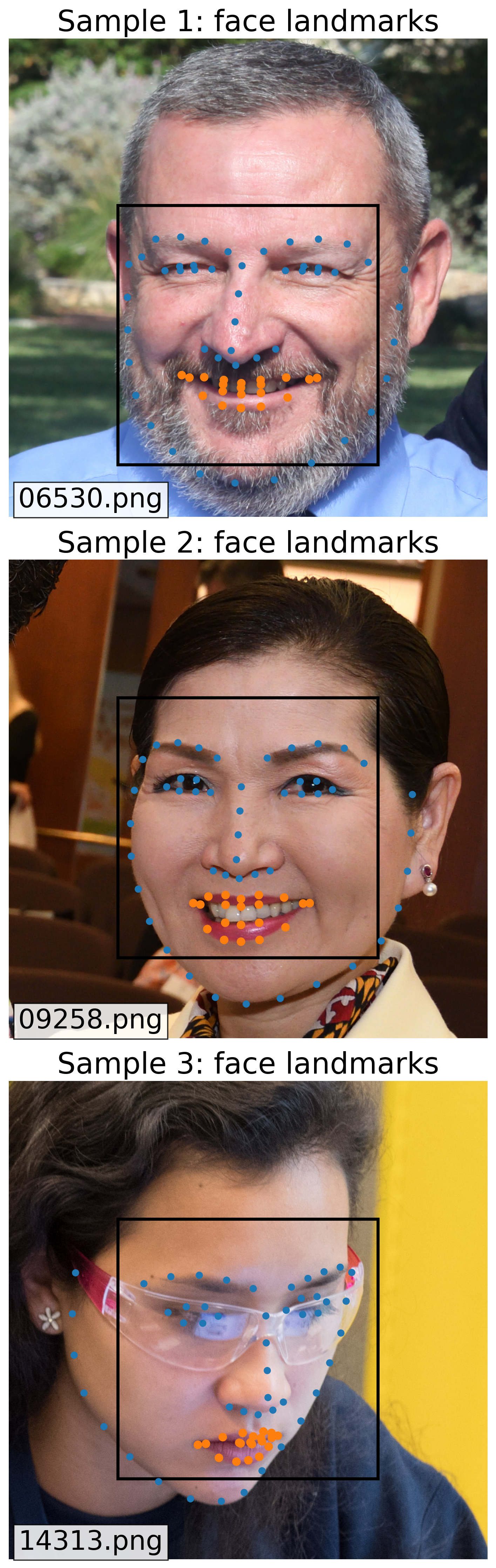}%
                }
                \caption{Face landmarks}
                \label{fig:face_landmarks}
            \end{subfigure}
            \hspace{1em}
            \begin{subfigure}[t]{0.491\columnwidth}
                \centering
                \includegraphics[width=\linewidth]{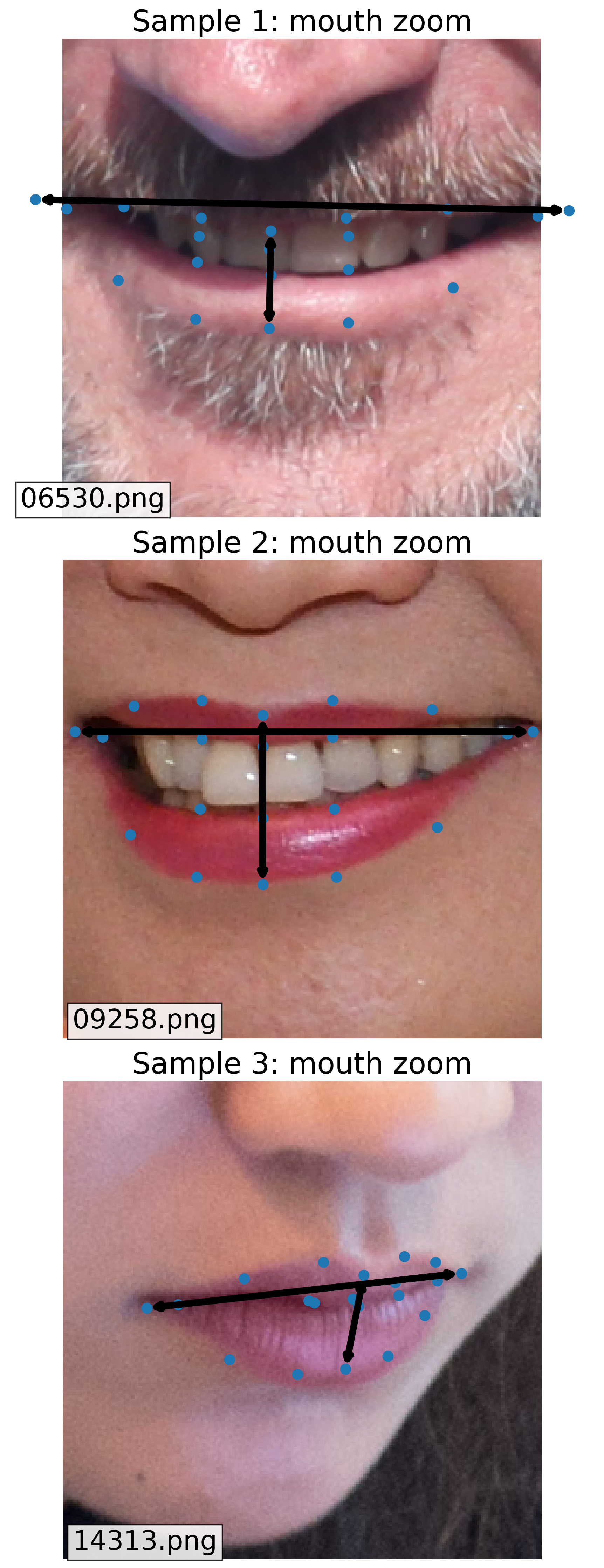}
                \caption{Mouth region zoom}
                \label{fig:mouth_zoom}
            \end{subfigure}
        \end{minipage}
    \end{adjustbox}
    \caption{Morphological detection stage of the \textit{AVSynthGen} pipeline. Facial landmarks are extracted using dlib \citep{king2009dlib}, and the mouth region is defined from the corresponding keypoints. Inter-landmark distances are used to evaluate mouth visibility through geometric thresholds.}
    \label{fig:landmark_pipeline}
\end{figure}

    The pipeline begins with a dataset of face images covering a diverse set of individuals, encompassing variability in gender, ethnicity, camera viewpoints, and age. These images are processed using a face detection and landmark-based morphological filtering stage specifically designed as part of the proposed \textit{AVSynthGen} pipeline and implemented on top of the dlib library \citep{king2009dlib}. First, a frontal face detector is applied to localize facial regions, followed by the extraction of 68 facial landmarks, whose structure is illustrated in Figure~\ref{fig:landmark_template}. Based on these landmarks, we define the mouth region by selecting the subset of points corresponding to the lip contour and inner mouth area (as illustrated in Figure~\ref{fig:landmark_pipeline}) and its visibility is assessed through simple geometric criteria derived from mouth width and opening estimated from inter-landmark distances. Images that do not satisfy these mouth visibility constraints are discarded, ensuring a minimum level of facial detail required for reliable lip motion synthesis.

\begin{figure}[pos=t]
      \centering
      \setlength{\tabcolsep}{1.5pt}
      \renewcommand{\arraystretch}{0.9}
    
      \begin{tabular}{@{}cccc@{}}
        \includegraphics[width=0.23\linewidth]{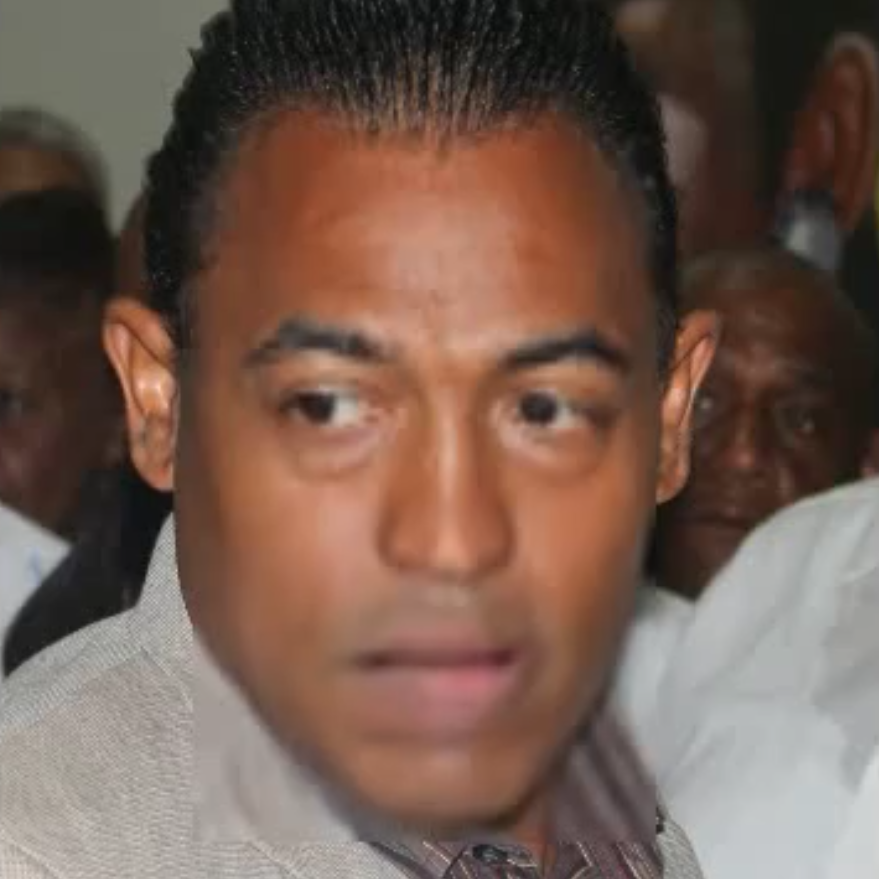} &
        \includegraphics[width=0.23\linewidth]{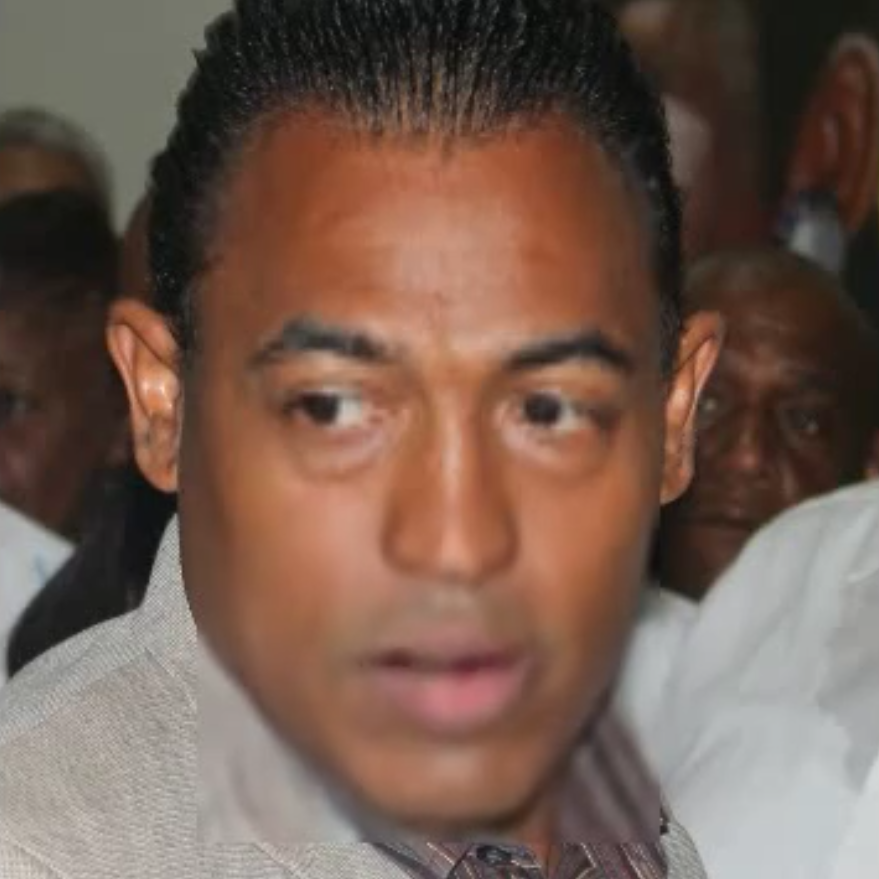} &
        \includegraphics[width=0.23\linewidth]{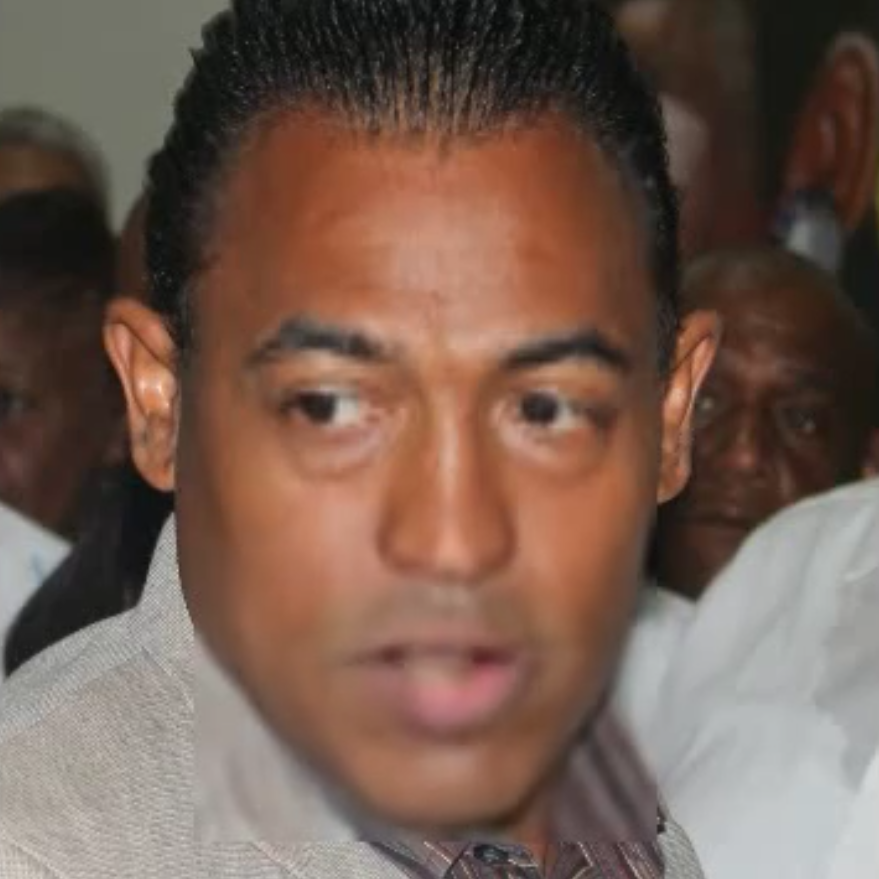} &
        \includegraphics[width=0.23\linewidth]{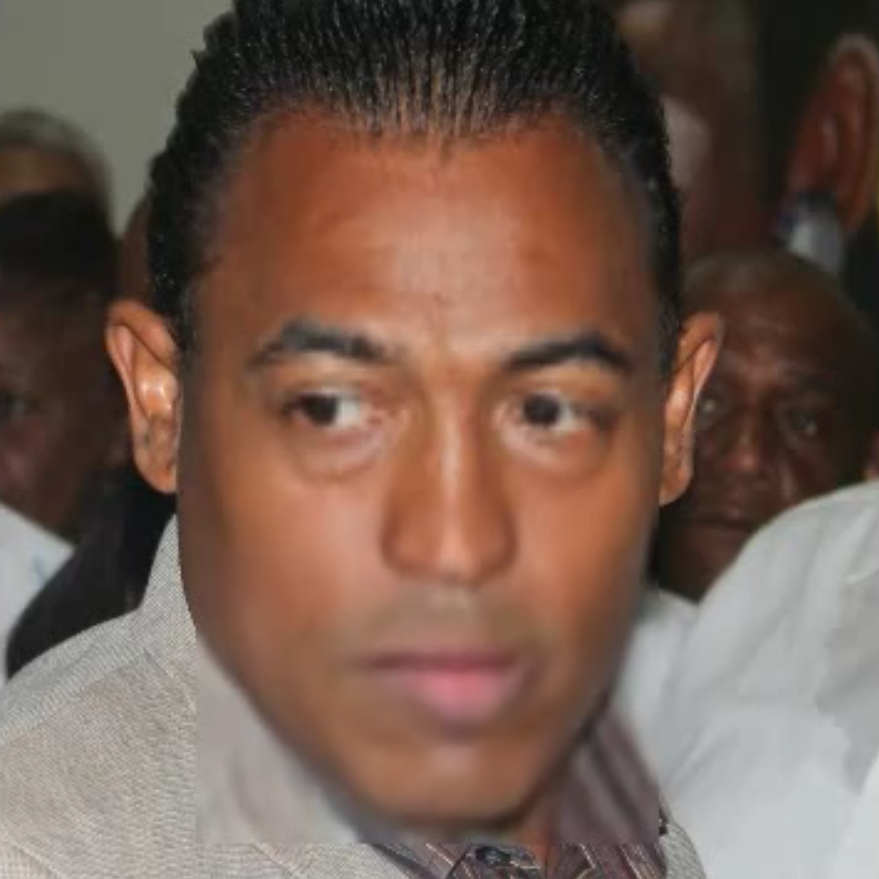} \\[0.5pt]
        \includegraphics[width=0.23\linewidth]{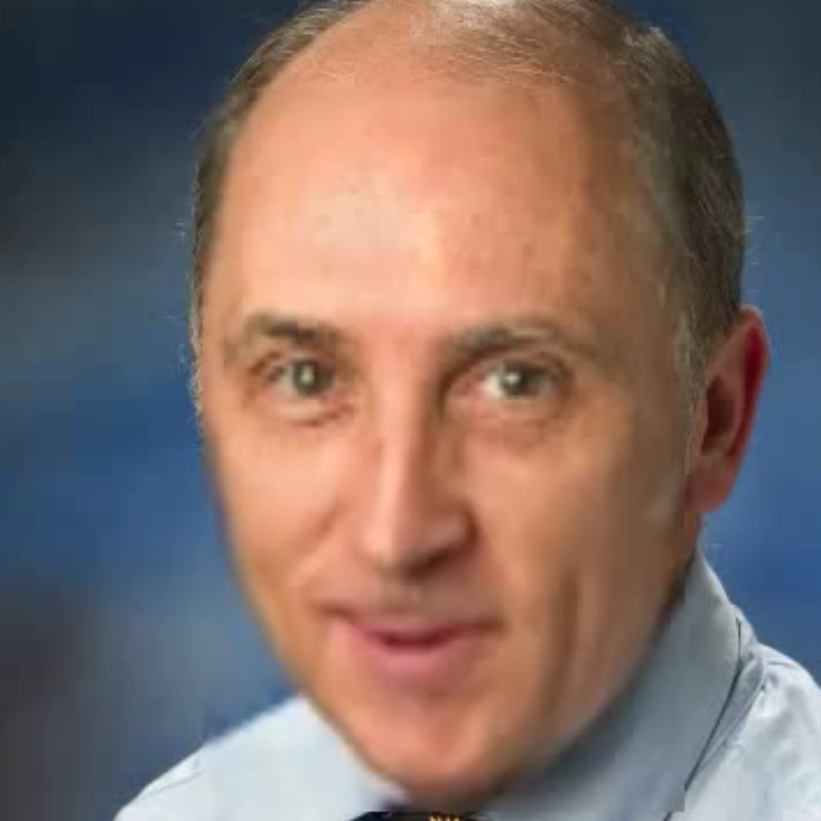} &
        \includegraphics[width=0.23\linewidth]{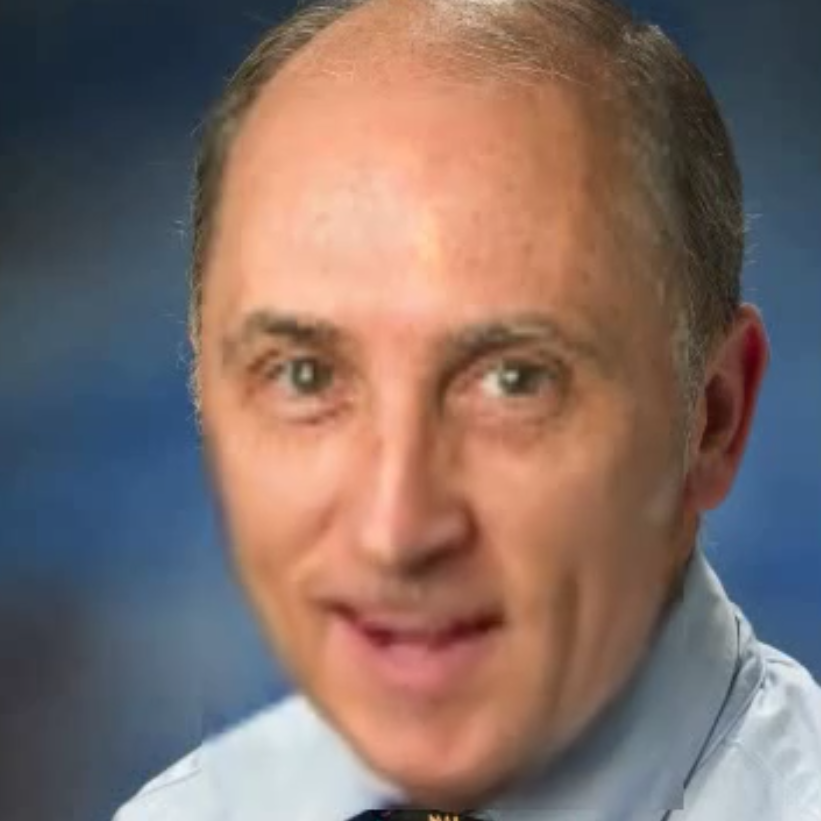} &
        \includegraphics[width=0.23\linewidth]{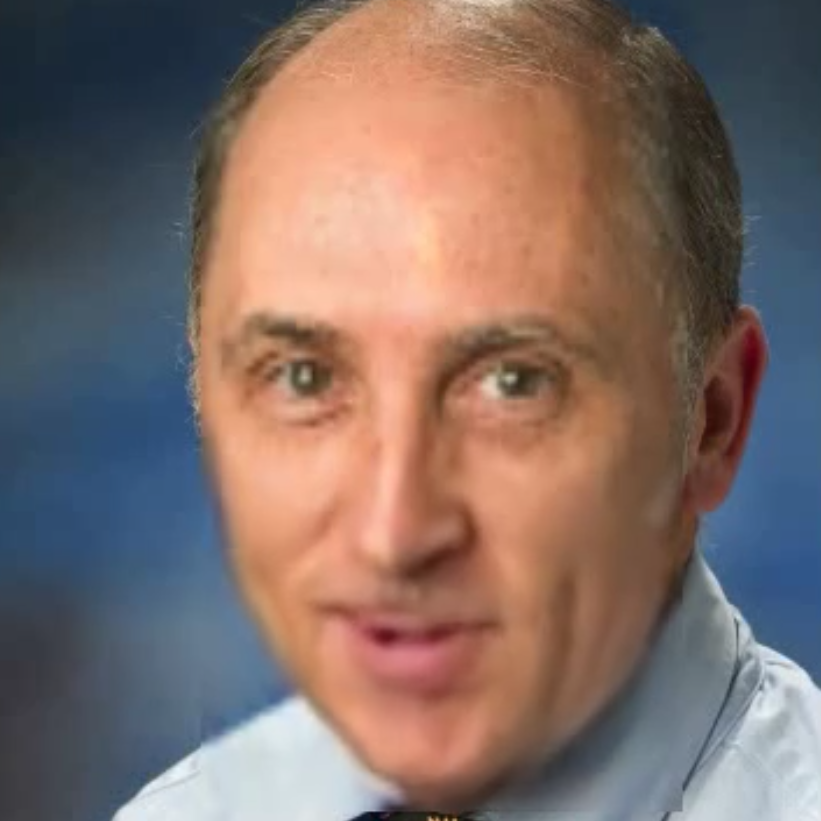} &
        \includegraphics[width=0.23\linewidth]{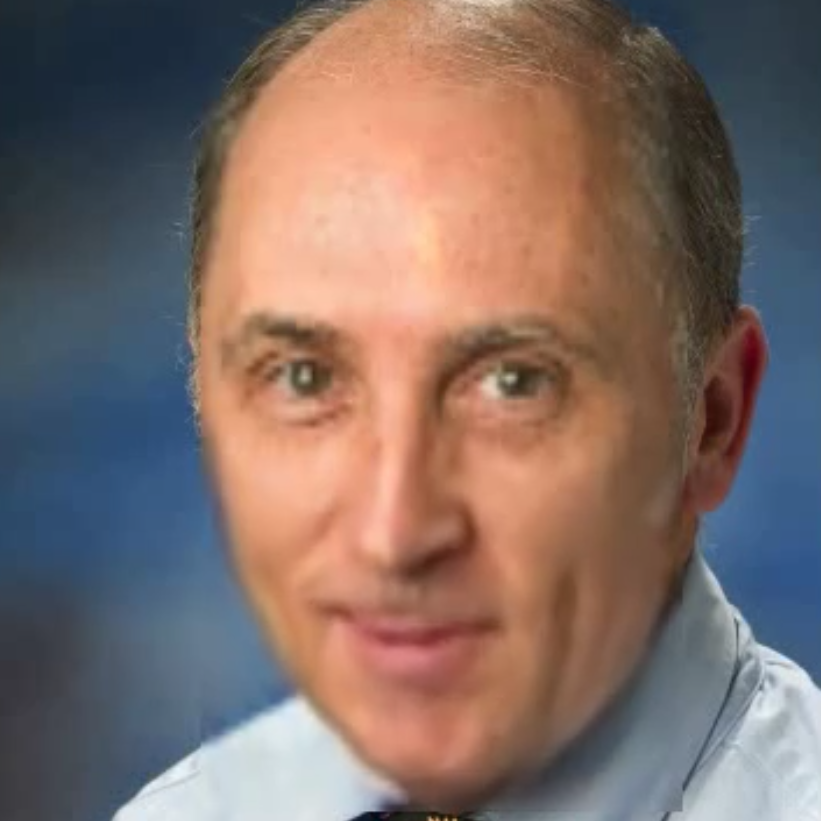} \\[0.5pt]
        \includegraphics[width=0.23\linewidth]{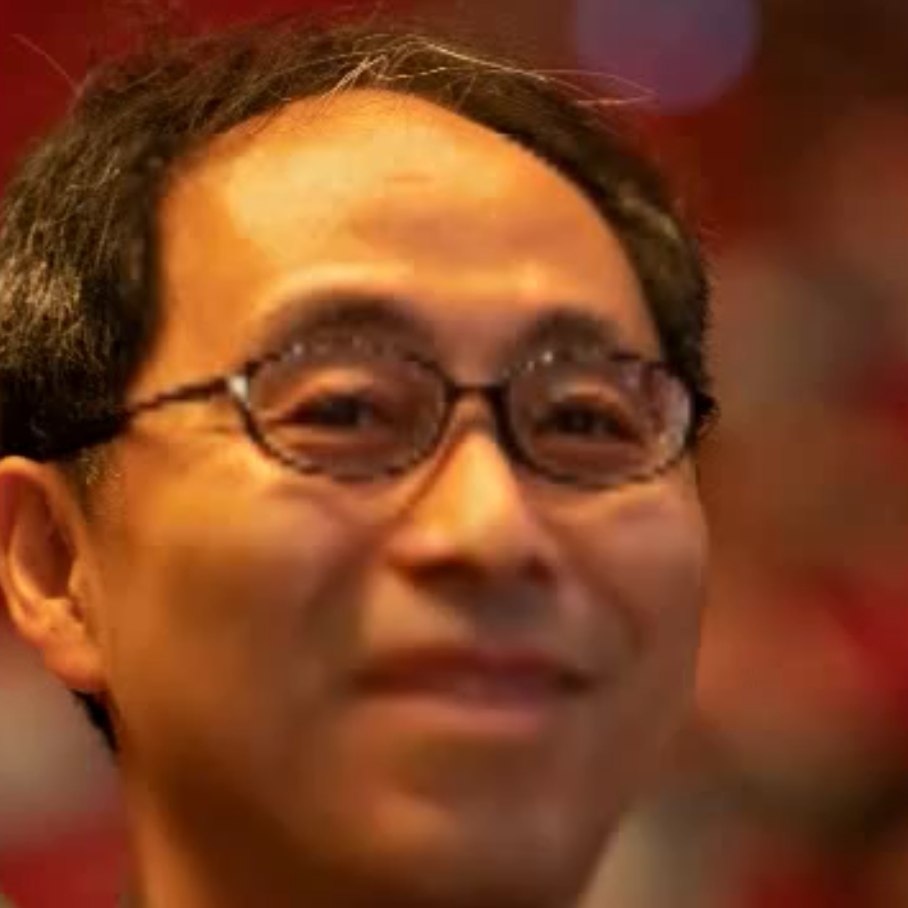} &
        \includegraphics[width=0.23\linewidth]{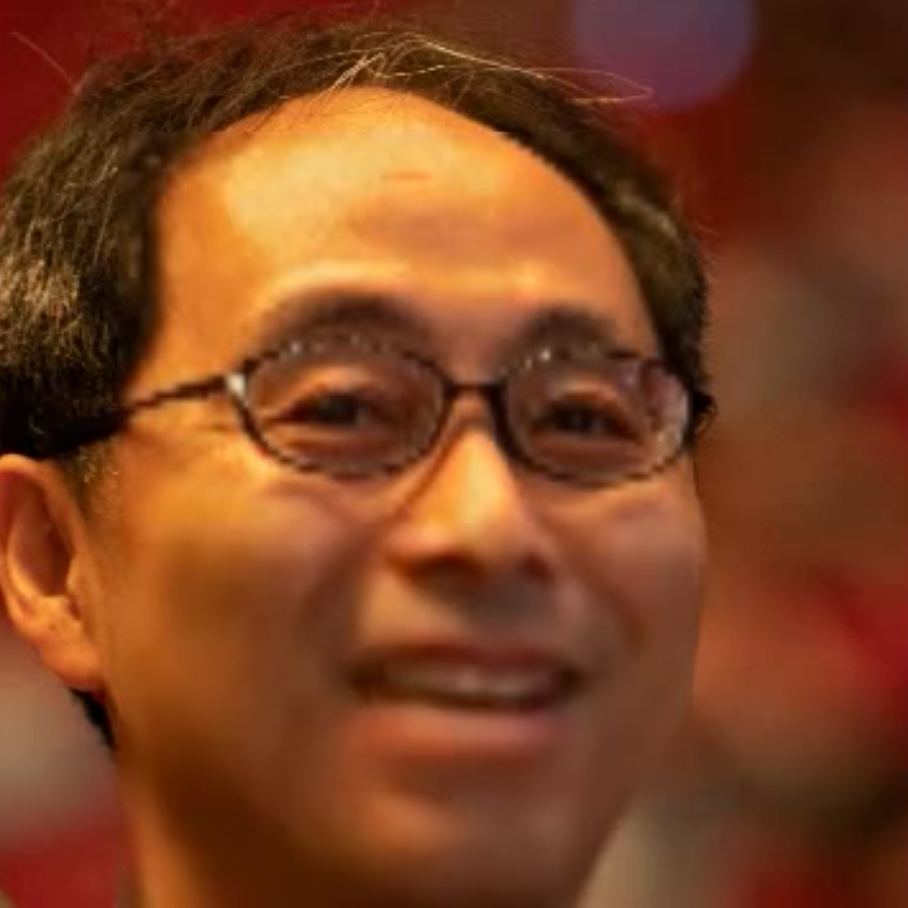} &
        \includegraphics[width=0.23\linewidth]{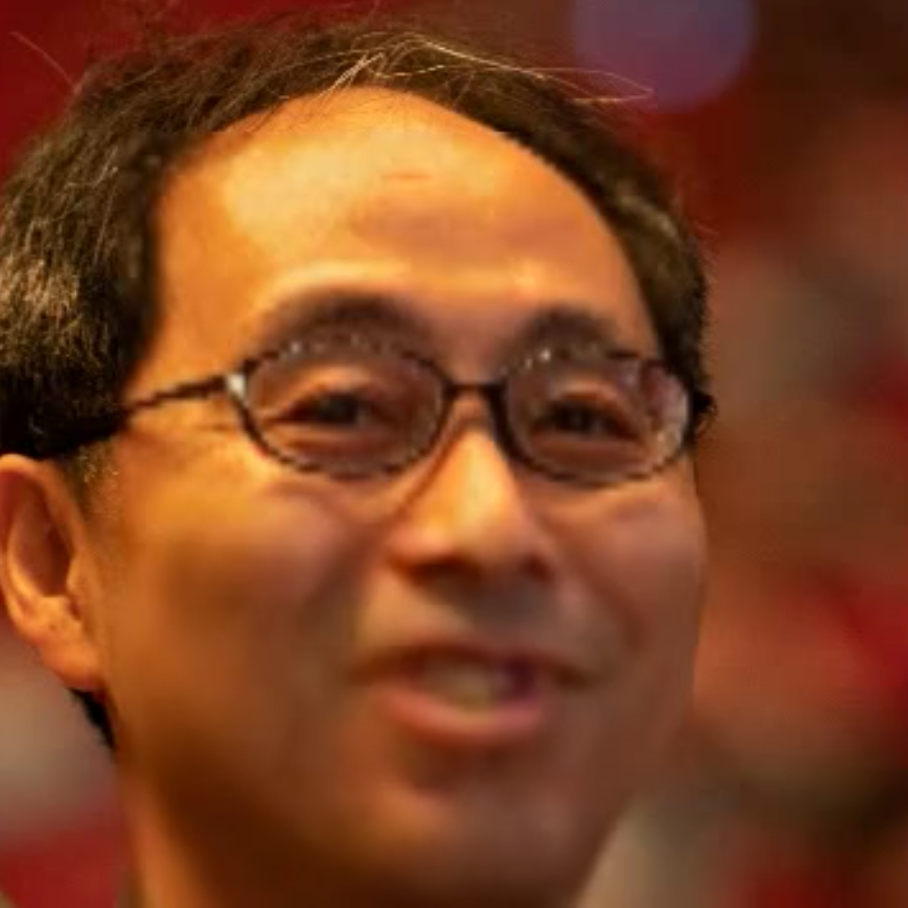} &
        \includegraphics[width=0.23\linewidth]{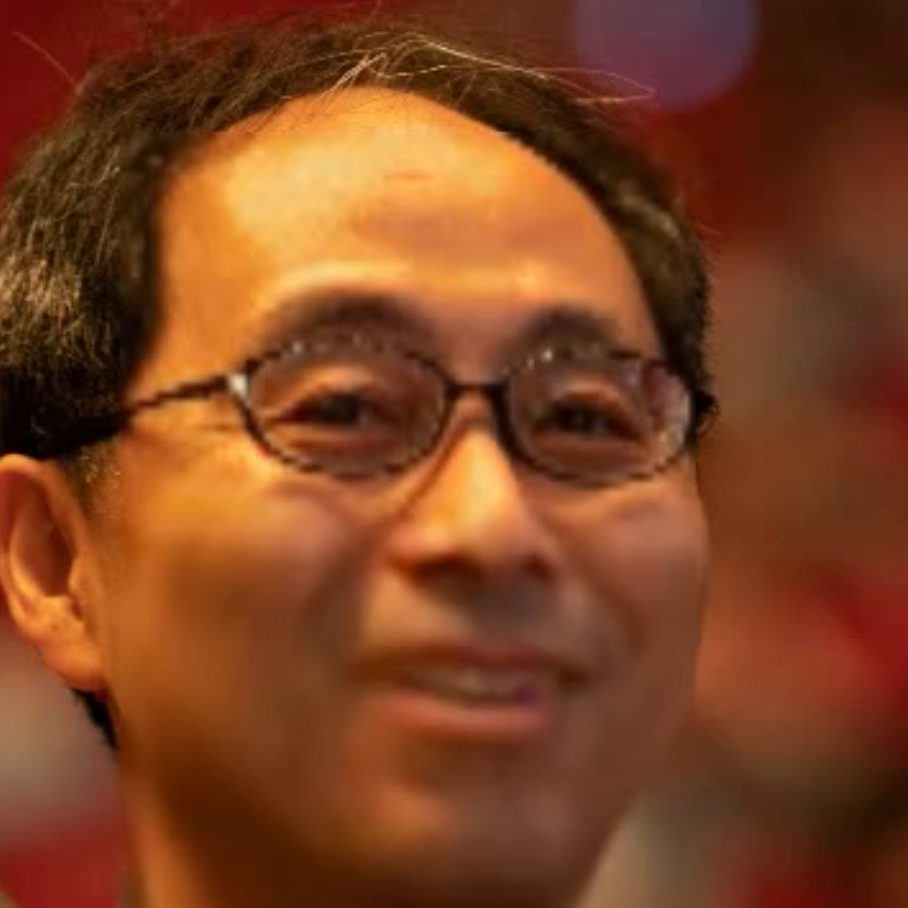}
      \end{tabular}
    
      \caption{Synthetic visual examples generated by the \textit{AVSynthGen} pipeline. The images show lip-synchronized faces produced by animating static images using real speech audio.}
      \label{fig:gan_only_examples}
    \end{figure}

    In parallel, an ASR dataset provides speech recordings and their corresponding transcriptions. For each audio sample, a face image is randomly selected from the previously filtered image set and used to generate a still-frame video by repeating the same image over the duration of the audio, resulting in temporally aligned image sequences that serve as input to the animation stage. These still-frame videos, together with the corresponding speech signals, are then processed by a lip synchronization model acting as a talking head generator focused specifically on the mouth region. Rather than synthesizing full facial motion, the model modifies only the lip area to match the input speech while preserving the rest of the face unchanged, transforming static facial images into temporally consistent visual speech sequences and producing synchronized audiovisual samples, as illustrated in Figure~\ref{fig:gan_only_examples}.

    Talking head generation models enable speech-driven facial animation directly from still images, removing the need for preexisting video data \citep{zhen2023talkingheadsurvey}. This choice is motivated by two main factors. First, constructing large-scale audiovisual datasets with sufficient diversity remains challenging and would reintroduce the AV data scarcity problem this work aims to address. Second, image datasets naturally provide a broad range of poses and camera viewpoints without requiring temporally continuous recordings. As a result, the pipeline can leverage widely available image and audio resources to generate diverse audiovisual data.
    
    The animation stage is implemented using the Wav2Lip + GAN model \citep{Prajwal2020Wav2Lip}, a state-of-the-art system for speech-driven facial animation that produces realistic lip movements aligned with a given audio input. We employ an enhanced variant of Wav2Lip that integrates a generative adversarial network (GAN) \citep{goodfellow2014gan} to improve the visual quality and realism of the synthesized mouth movements. This model ensures accurate temporal alignment between audio and visual streams while maintaining perceptual consistency in the generated facial dynamics.
    
    The final output of the pipeline is a synthetic audiovisual dataset composed of generated video sequences paired with transcriptions derived from the original speech corpus. The AVSynthGen framework enables scalable data generation from existing resources, facilitating the creation of large multimodal datasets for languages with limited audiovisual data availability\footnote{Code for the \textit{AVSynthGen} pipeline is available at \url{https://github.com/Pol-Buitrago/SynthAVSR}}.

\subsection{Semi-automatic Annotation Pipeline}
\label{subsec:annotation}
    
    As discussed in Section~\ref{sec:introduction}, no labeled Catalan audiovisual corpus is currently available, which motivates our synthetic data generation strategy. However, while this approach still allows us to train an AVSR model for zero-AV-resource languages such as Catalan, it does not solve the evaluation problem, since the absence of real AV data still prevents a direct assessment of the resulting system. To enable meaningful evaluation in this zero-AV-resource setting, we additionally developed a semi-automatic annotation pipeline for constructing a Catalan AV test set from raw, unlabeled video material\footnote{Code available in the same repository as \textit{AVSynthGen}, see Section~\ref{subsec:pipeline}.} and use it to evaluate our resulting AVSR models.

\begin{figure*}[t]
        \centering
        \includegraphics[width=\linewidth]{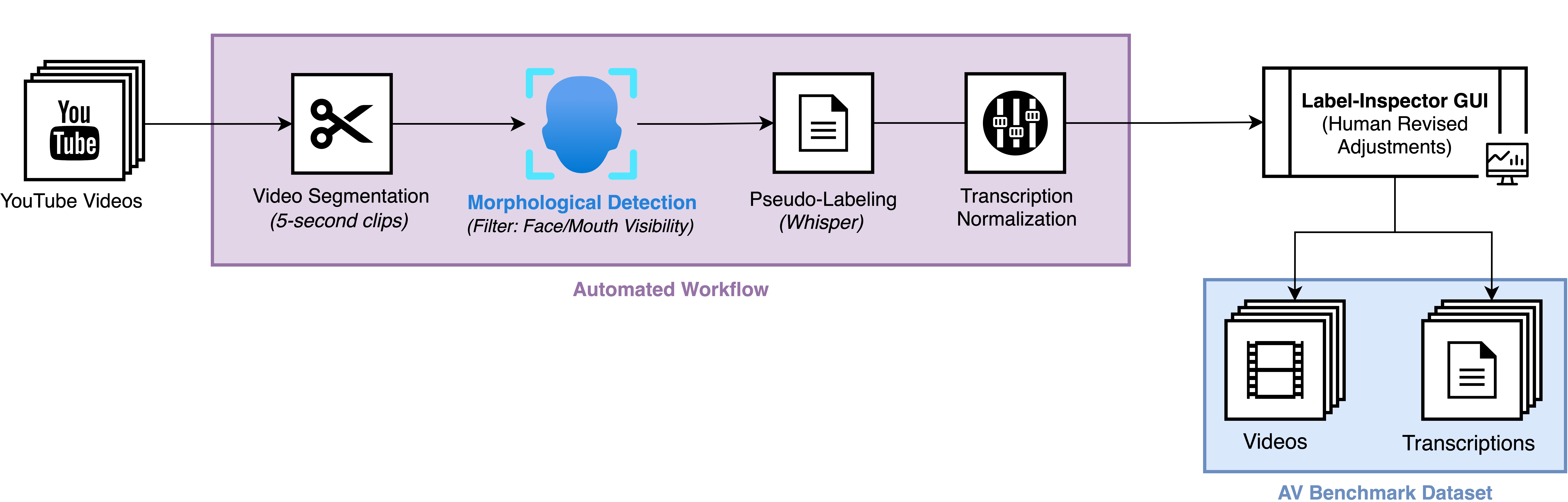}
        \caption{Semi-automatic pipeline for annotating the Catalan AV benchmark. Raw video material is segmented into short clips, filtered using the morphological detection stage (Section~\ref{subsec:pipeline}) to ensure mouth visibility, automatically pseudo-labeled with an ASR system \citep{radford2022whisper}, normalized, and manually reviewed and corrected via a custom graphical interface (\textit{Label-Inspector}).}
        \label{fig:annotation_pipeline}
    \end{figure*}

    The annotation pipeline operates by segmenting raw video material into short clips, which are first filtered using the landmark-based morphological filtering stage described in Section~\ref{subsec:pipeline} and illustrated in Figures~\ref{fig:landmark_template} and~\ref{fig:landmark_pipeline}, ensuring clear facial and mouth visibility in all retained samples. The filtered segments are then automatically pseudo-labeled using a state-of-the-art ASR system \citep{radford2022whisper}, after which transcriptions are normalized. The resulting labels are manually reviewed and corrected using a developed custom graphical interface, which we refer to as \textbf{\textit{Label-Inspector}}, enabling accurate human annotation with minimal manual effort. Figure~\ref{fig:annotation_pipeline} summarizes the annotation workflow.

    This tool was developed primarily to obtain a reliable evaluation benchmark for the experiments in Catalan presented in this work, but it can naturally be extended into a general-purpose AV annotation tool for low-resource languages, although this lies beyond the scope of the present work.

\subsection{AVSR Model and Training}
    We adopt the AV-HuBERT framework introduced by \citet{shi2022avhubert} and extended in \citet{shi2022avsr} as the baseline in our experiments. AV-HuBERT is an open-source, well-established self-supervised architecture for learning joint audio-visual representations and has demonstrated strong performance across multiple AVSR benchmarks.

    Rather than training from scratch, our strategy focuses on fine-tuning AV-HuBERT to our specific experimental setup, leveraging publicly available pre-trained checkpoints designed to be generic and adaptable to multiple tasks. 
    The selected checkpoint\footnote{Referenced checkpoint: \url{https://dl.fbaipublicfiles.com/avhubert/model/lrs3_vox/clean-pretrain/large_vox_iter5.pt}} was pre-trained on unlabeled audiovisual data from the English LRS3 \citep{afouras2018lrs3ted} and VoxCeleb2 \citep{chung2018voxceleb2} datasets. Although these corpora contain only English speech, prior research on multilingual pre-training of language models has shown significant transfer capabilities to other languages \citep{conneau2020xlmr}, suggesting that similar benefits can be leveraged for AVSR in low-resource language scenarios.
        
    For our setting, we adopt an attention-based sequence-to-sequence fine-tuning strategy with cross-entropy loss \citep{bahdanau2016endtoend}, following the formulation described by \citet{shi2022avhubert}. A randomly initialized 6-layer Transformer decoder is attached to the pre-trained encoder to map its output representations into SentencePiece (unigram) subword units \citep{kudo2018subword}, enabling robust recognition across our target languages.

    The model is optimized using the Adam optimizer \citep{kingma2015adam} with a base learning rate of $1\times10^{-3}$ and a tri-stage learning rate schedule with warmup and decay. To stabilize training, the pre-trained encoder is partially frozen during the initial phase (first 22,500 updates), after which full fine-tuning is performed. Model selection is based on validation performance.
        
\section{Experimental Setup}\label{sec:setup}
\subsection{Datasets}\label{subsec:datasets}

    The dataset collection is divided into two categories: first, the datasets used to generate synthetic audiovisual data through the pipeline described in Section~\ref{subsec:pipeline}, consisting of audio and image sources; and second, the real audiovisual datasets used for training and evaluation.

    \subsubsection{Synthetic Data Generation}   
    
        For the image dataset, we employed the Flickr-Faces-HQ (FFHQ) dataset \citep{Karras2019StyleGAN}, which provides high-resolution images of human faces with significant variation in facial attributes, poses, and identities, making it particularly well-suited for our lip synchronization task. 
        
        For Spanish audio, we used the Spanish subset of the Mozilla Common Voice dataset (version 19.0) \citep{ardila2020commonvoice} ($\approx$512 hours), a large-scale corpus of manually validated speech recordings.
        For Catalan audio, we relied on two primary sources: TV3Parla \citep{kulebi18_iberspeech} ($\approx$291 hours) and ParlamentParla \citep{kulebi2022parlamentparla} ($\approx$432 hours). TV3Parla consists of broadcast speech from television programs, while ParlamentParla contains recordings from parliamentary sessions, providing complementary variations in speaking style, formality, and acoustic conditions.
    
        It is important to note that we excluded the test splits from all audio datasets, as the audio data is used exclusively for the generation of synthetic audiovisual samples. Evaluation is conducted only on real audiovisual data to ensure a realistic assessment of model performance.
    
        The selected audio data is then paired with the previously described image samples to generate synthetic audiovisual datasets, following the speech-driven visual synthesis procedure detailed in Section~\ref{subsec:pipeline}. For clarity, we refer to the resulting audiovisual versions of each audio corpus as \textbf{SynthAV-CV} (synthetic audiovisual version of Common Voice), \textbf{TV3ParlAV} (synthetic audiovisual version of TV3Parla), and \textbf{ParlAVment} (synthetic audiovisual version of ParlamentParla). A summary of each corpus is provided in Table~\ref{tab:av_datasets}.
    
    \subsubsection{Real Audiovisual Data}
        Following the two experimental scenarios described in Section~\ref{sec:introduction}, Spanish is used as a low-resource augmentation setting, where synthetic data is used to complement existing real audiovisual corpora, while Catalan is used as a zero-AV-resource setting, where training relies exclusively on synthetic audiovisual data.

        For this purpose, in the Spanish setting, we employ three widely used Spanish audiovisual speech corpora: \textbf{LIP-RTVE} \citep{lleida2019albayzin, liprtve2022lrec} ($\approx$13h), \textbf{CMU-MOSEAS\textsubscript{es}} \citep{bagherzadeh2020cmumoseas} ($\approx$13h), and \textbf{MuAViC\textsubscript{es}} \citep{anwar2023muavic} ($\approx$180h) (see Table~\ref{tab:av_datasets}). The train and validation subsets are used for model training, while the test subsets are used to evaluate model performance on real audiovisual data.

        Although the Catalan experiments are conducted under a zero-AV-resource setting, real audiovisual data is still required for meaningful evaluation. However, no existing Catalan audiovisual corpus is available. To address this, we used the proposed annotation pipeline described in Section~\ref{subsec:annotation} to construct a manually curated test set for Catalan evaluation.

        The dataset, named \textbf{AV-CAT}, consists of 51 minutes and 38 seconds of footage (see Table~\ref{tab:av_datasets}), divided into approximately 5-second segments extracted from videos of the public Catalan television channel. These segments ensure clear visibility of the mouth and adhere to diversity criteria in content. AV-CAT serves as the evaluation benchmark for Catalan, providing the first labeled AV dataset for this language.

    \begin{table}[t]
        \centering
        \caption{AV datasets employed in our experiments.}
        \label{tab:av_datasets}
    
        \small
        \setlength{\tabcolsep}{4pt}
    
        \begin{tabular}{@{}l@{\hspace{30pt}}c@{\hspace{15pt}}l@{\hspace{5pt}}}
        \toprule
        \textbf{Name} & \textbf{Language} & \textbf{Duration} \\ 
        \midrule
    
        \multicolumn{3}{@{}l@{}}{\textbf{--- Real AV Datasets ---}} \\
        LIP-RTVE & Spanish & $\approx$13h \\
        CMU-MOSEAS\textsubscript{es} & Spanish & $\approx$13h \\
        MuAViC\textsubscript{es} & Spanish & $\approx$180h \\
        AV-CAT (custom test) & Catalan & $\approx$51min \\
    
        \addlinespace
    
        \multicolumn{3}{@{}l@{}}{\textbf{--- Synthetic AV Datasets ---}} \\
        SynthAV-CV & Spanish & $\approx$512h \\
        TV3ParlAV & Catalan & $\approx$291h \\
        ParlAVment & Catalan & $\approx$432h \\ 
    
        \bottomrule
        \end{tabular}
    \end{table}

    As shown in Table~\ref{tab:av_datasets}, the synthetic datasets have a considerably longer duration than the real AV datasets. This difference reflects the inherent scalability advantage of synthetic data. Real AV corpora are constrained by the cost of audiovisual collection and annotation, whereas synthetic visual streams can be generated from large ASR resources without such limitations.
    
    Although matching dataset sizes could provide a more controlled comparison, it would eliminate one of the main practical advantages of synthetic data, namely its scalability. Rather than demonstrating that synthetic visual data is equivalent or superior to real audiovisual data under strictly controlled conditions, our objective is to evaluate whether it can serve as an effective and scalable alternative when real AV resources are limited or unavailable.
    
\subsection{Experiments}
\label{subsec:experiments}
    This section describes the models trained and the experimental protocol designed to measure the contribution of synthetic visual data to audiovisual speech recognition. We evaluate two language settings: Spanish (limited but available AV resources) and Catalan (a zero-AV-resource scenario). Word Error Rate (WER) is used as the sole evaluation metric, with model selection based on validation WER.

    \subsubsection{Training configurations}
    Our training process follows a comparative approach by evaluating three audiovisual training configurations that differ only in the source of the visual modality. Since the audio stream is always real, the configuration names refer exclusively to the origin of the \emph{visual} training data. Specifically, we compare training with exclusively real visual data, exclusively synthetic visual data, and a combination of both. This design allows us to assess the impact of synthetic visual content on AVSR performance and its potential to complement or replace real-world audiovisual datasets.

    \begin{table}[t]
        \caption{Overview of the training datasets used for each AVSR training configuration. Configuration names indicate the source of the visual training data and the target language.}
        \label{tab:trained_models}
        \setlength{\tabcolsep}{3pt}
        \fontsize{8}{10}\selectfont
        \begin{tabularx}{\columnwidth}{lX}
        \toprule \\[-2.5ex]
            \textbf{Configuration} & \textbf{Training Datasets} \\\\[-2ex]
            \hline \\[-1ex]
            \textit{RealVisual\textsubscript{es}} & {LIP-RTVE, CMU-MOSEAS\textsubscript{es}, MuAViC\textsubscript{es} \textbf{($\approx$206h)}} \\\\[-1ex]
            \textit{SynthVisual\textsubscript{es}} & {SynthAV-CV \textbf{($\approx$512h)}} \\\\[-1ex]
            \textit{MixedVisual\textsubscript{es}} & {LIP-RTVE, CMU-MOSEAS\textsubscript{es}, MuAViC\textsubscript{es}, SynthAV-CV \textbf{($\approx$718h)}} \\
            \multicolumn{2}{@{}l@{}}{\hdashrule{\linewidth}{0.4pt}{1mm 1mm}} \\\\[-2ex]
            \textit{SynthVisual\textsubscript{cat}} & {TV3ParlAV, ParlAVment \textbf{($\approx$723h)}} \\[1ex]
            \bottomrule
        \end{tabularx}
    \end{table}

    Under the Spanish setting, we define three AVSR training configurations: \textit{\textbf{RealVisual\textsubscript{es}}}, trained using real AV data; \textit{\textbf{SynthVisual\textsubscript{es}}}, trained using synthetic AV data generated from real speech recordings; and \textit{\textbf{MixedVisual\textsubscript{es}}}, trained on the union of both datasets, combining all real AV samples from \textit{RealVisual\textsubscript{es}} with the complete synthetic AV corpus from \textit{SynthVisual\textsubscript{es}}. Under the Catalan setting, given the absence of real audiovisual datasets, we define a single training configuration, \textit{\textbf{SynthVisual\textsubscript{cat}}}, using only synthetic audiovisual data. The training datasets corresponding to each configuration are summarized in Table~\ref{tab:trained_models}.

    Each training configuration is used to train three model variants using the same training dataset but differing only in the input modalities available during training: an audiovisual (AV) model, which has access to both audio and video; an audio-only (A) model, in which the visual input is masked; and a video-only (V) model, in which the audio input is masked. This design enables a controlled evaluation of the contribution of each modality.

\subsubsection{Experiments in Spanish}
    The experiments in Spanish are designed to evaluate the three training configurations introduced above, \textit{RealVisual\textsubscript{es}}, \textit{SynthVisual\textsubscript{es}}, and \textit{MixedVisual\textsubscript{es}}, and to assess the impact of synthetic visual data on audiovisual speech recognition. As described previously, each training configuration is used to train three modality-specific variants: audiovisual (AV), audio-only (A), and video-only (V).
    
    Since the audio stream is always real, evaluating only the audiovisual models would make it difficult to isolate the contribution of synthetic visual data, as it could be masked by the audio modality. Therefore, we first evaluate the video-only (V) variants to determine whether synthetic visual data alone provides meaningful articulatory information for speech recognition. This initial analysis determines whether the synthesized visual stream contributes useful visual cues before assessing its impact in the multimodal setting.
    
    Once the usefulness of the synthetic visual modality has been established, we evaluate the audiovisual (AV) variants on the same benchmarks to quantify the effectiveness of synthetic visual data as an augmentation strategy and compare our best-performing model against previously reported AVSR baselines.
    
    Beyond the primary performance evaluation, we perform a series of complementary analyses. First, we compare the AV, A, and V variants of the best-performing training configuration under identical experimental conditions to quantify the contribution of each modality. Next, we investigate modality reliance using Integrated Gradients (IG) \citep{sundararajan2017integratedgradients} by comparing the video-channel attributions of the AV variants of \textit{RealVisual\textsubscript{es}} and \textit{SynthVisual\textsubscript{es}}. Finally, we assess robustness under adverse acoustic conditions to determine whether the multimodal benefits obtained through synthetic visual training are preserved in noisy environments.

\subsubsection{Experiments in Catalan}
    The Catalan experimental setting represents a fundamentally different scenario from the Spanish setting. Since no labeled audiovisual datasets are available for Catalan, it constitutes a zero-AV-resource setting in which training relies exclusively on synthetic audiovisual data. In this scenario, we evaluate the audiovisual (AV), audio-only (A), and video-only (V) variants of \textit{SynthVisual\textsubscript{cat}} on our manually annotated Catalan AV test set (Section~\ref{subsec:datasets}). 
    
    The objective is to determine whether synthetic audiovisual data alone can train an effective AVSR model and, in particular, whether multimodal training provides measurable improvements over an audio-only system despite the absence of real audiovisual training data.

\section{Results}
    This section presents the results of the experiments described in Section~\ref{subsec:experiments}, analyzing the impact of synthetic visual data across different evaluation settings.
    
    Regarding the Spanish experiments, Section~\ref{subsec:lipreading_performance} reports visual-only speech recognition performance with synthetic video, Section~\ref{subsec:avsr_performance} reports AVSR performance, including comparisons with published models trained under the same baseline, and Section~\ref{subsec:unimodal_vs_multimodals} compares AVSR against unimodal baselines. Section~\ref{subsec:input_attribution_analysis} then investigates modality reliance through input attribution methods, and Section~\ref{subsec:noise_robustness} evaluates robustness under varying noise conditions. 
    
    Finally, Section~\ref{subsec:zero_resource_language_adaptability} reports results for Catalan, testing whether synthetic audiovisual data alone can train a functional AVSR model in a zero-AV-resource scenario.

\subsection{VSR Performance with Synthetic Video} 
\label{subsec:lipreading_performance}
\begin{figure}[pos=t]
      \centering
      \resizebox{\linewidth}{!}{
        \begin{tikzpicture}
          \begin{axis}[
            ybar=2pt,
            symbolic x coords={LIP-RTVE, CMU-MOSEAS\textsubscript{es}, MuAViC\textsubscript{es}, Average},
            xtick=data,
            xticklabels={LIP-RTVE, CMU-MOSEAS\textsubscript{es}, MuAViC\textsubscript{es}, \textbf{Average}},
            ylabel={Word Error Rate (\%)},
            y label style={at={(axis description cs:-0.095,.5)}, anchor=south},
            bar width=15pt,
            enlarge x limits=0.2,
            ymin=65, ymax=115,
            grid=major,
            legend style={at={(0.43,0.95)}},
            height=8cm,
            width=1.25\linewidth,
            legend image code/.code={%
              \draw[#1] (0.1cm,-0.1cm) rectangle (0.5cm,0.1cm);
            }, 
            x tick label style={rotate=25, yshift=5pt},
          ]
    
          % RealVisual
          \addplot[
            fill={rgb, 1:red, 0.85; green, 0.85; blue, 0.85},
            pattern=north west lines,
          ] coordinates {(LIP-RTVE, 90.4) (CMU-MOSEAS\textsubscript{es}, 90.7) (MuAViC\textsubscript{es}, 102) (Average, 94.37)};
    
          % SynthVisual
          \addplot[
            fill={rgb, 1:red, 0.7; green, 0.7; blue, 0.7},
            pattern=crosshatch,
          ] coordinates {(LIP-RTVE, 97.4) (CMU-MOSEAS\textsubscript{es}, 97) (MuAViC\textsubscript{es}, 101.9) (Average, 98.77)};
    
          % MixedVisual
          \addplot[
            fill={rgb, 1:red, 0.3; green, 0.3; blue, 0.3},
          ] coordinates {(LIP-RTVE, 68) (CMU-MOSEAS\textsubscript{es}, 67.1) (MuAViC\textsubscript{es}, 98) (Average, 77.7)};
    
          \legend{RealVisual\textsubscript{es}, SynthVisual\textsubscript{es}, MixedVisual\textsubscript{es}}
    
          \end{axis}
        \end{tikzpicture}
      }
      \caption{WER comparison for the video-only modality on different datasets. The ``Average'' bar represents the mean WER across the three datasets.}
      \label{fig:videoonly_results}
    \end{figure}
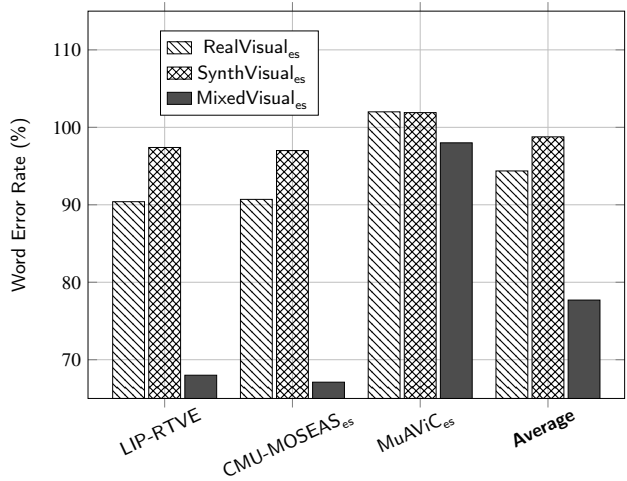

    As a first step, we evaluate whether synthetic visual data provides meaningful articulatory information on its own, isolating the visual modality from the audio stream. Table~\ref{tab:wer_results_V} and Figure~\ref{fig:videoonly_results} report the Visual Speech Recognition (VSR) performance of the three training configurations on different benchmarks.

    \begin{table}[pos=t]
        \centering
        \caption{WER comparison of the trained Spanish models in the video-only modality on different benchmarks. The values in parentheses indicate the relative improvement over the \textit{RealVisual\textsubscript{es}} (video-only) model.}
        \label{tab:wer_results_V}
        \setlength{\tabcolsep}{5.75pt}
        \fontsize{7}{10}\selectfont
        \begin{tabularx}{\columnwidth}{>{\raggedright\arraybackslash}X c c c}
            \toprule
              \textbf{Model} (V) & \textbf{LIP-RTVE} & \textbf{CMU-MOSEAS\textsubscript{es}} & \textbf{MuAViC\textsubscript{es}} \\
              \midrule
              \textit{RealVisual\textsubscript{es}}   & 90.4\% & 90.7\% & 102\% \\
              \textit{SynthVisual\textsubscript{es}}  & 97.4\% & 97\% & 101.9\% \\
              \multicolumn{4}{@{}l@{}}{\hdashrule{\linewidth}{0.4pt}{1mm 1mm}} \\
              \textit{MixedVisual\textsubscript{es}}    & \textbf{68\%{\fontsize{6}{10}\selectfont($\downarrow$24.7\%)}} & \textbf{67.1\%{\fontsize{6}{10}\selectfont($\downarrow$26\%)}} & \textbf{98\%{\fontsize{6}{10}\selectfont($\downarrow$3.9\%)}} \\
              \bottomrule
        \end{tabularx}
    \end{table}
    
    As shown in Table~\ref{tab:wer_results_V}, models trained exclusively on either real or synthetic visual data achieve limited visual speech recognition performance across all evaluated benchmarks. This outcome is expected for two main reasons. First, visual speech recognition is inherently more challenging than audio-based recognition because lip movements alone provide only partial information about the spoken content \citep{Sumby1954VisualCT,erber1975auditoryvisual}. Second, each training configuration presents its own limitations. The \textit{RealVisual\textsubscript{es}} model is trained on a relatively small amount of real audiovisual data, limiting the diversity of visual speech patterns it can learn. Conversely, although \textit{SynthVisual\textsubscript{es}} is trained on substantially more data, the synthesized visual streams remain less intelligible than real video, as also observed in previous work on synthetic lipreading \citep{shan2022speechinnoise,liu2023synthvsr}, where video-only WERs exceeding 100\% have also been reported.

    However, when synthetic video is used to augment the real audiovisual corpus, the \textit{MixedVisual\textsubscript{es}} model achieves substantially lower WERs on two of the three benchmarks (LIP-RTVE and CMU-MOSEAS\textsubscript{es}), with relative reductions of 24.7\% and 26.0\%, respectively. On MuAViC\textsubscript{es}, the improvement is smaller (3.9\%), which is expected given the greater variability and less controlled recording conditions of this corpus, where visual-only speech recognition is inherently more difficult.

    These results reveal a clear complementary effect between real and synthetic visual data. Neither source of visual data is sufficient on its own. Real audiovisual data provides high-fidelity visual speech examples but is limited in scale, whereas synthetic data substantially increases the amount and diversity of visual training samples despite its lower visual quality. Combining both sources allows the model to benefit simultaneously from the realism of the former and the scale of the latter, resulting in consistently improved VSR performance.
        
    \subsection{AVSR Performance with Synthetic Video} 
    \label{subsec:avsr_performance}
    The previous results demonstrate that synthetic visual data contributes meaningful articulatory information and can strengthen visual speech representations when used to augment real audiovisual data. 
    
    Once the effectiveness of synthetic visual data has been established in the video-only setting, we evaluate whether these benefits translate to audiovisual speech recognition. Specifically, we assess whether augmenting real AV training data with synthetic visual data improves multimodal speech recognition performance.

    \begin{table}[t]
    \centering
        \caption{WER comparison of the trained Spanish AVSR models on different benchmarks. The values in parentheses indicate the relative improvement over the \textit{RealVisual\textsubscript{es}} model.}
        \label{tab:wer_results}
        \setlength{\tabcolsep}{5pt}
        \fontsize{7}{10}\selectfont
        \begin{tabularx}{\columnwidth}{>{\raggedright\arraybackslash}X c c c}
            \toprule
              \textbf{Model} (AV) & \textbf{LIP-RTVE} & \textbf{CMU-MOSEAS\textsubscript{es}} & \textbf{MuAViC\textsubscript{es}} \\
              \midrule
              \textit{RealVisual\textsubscript{es}}   & 9.3\% & 15.4\% & 16.6\% \\
              \textit{SynthVisual\textsubscript{es}}  & 21.1\% & 35.2\% & 39.6\% \\
              \multicolumn{4}{@{}l@{}}{\hdashrule{\linewidth}{0.4pt}{1mm 1mm}} \\
              \textit{MixedVisual\textsubscript{es}}    & \textbf{8.1\%{\fontsize{6}{10}\selectfont($\downarrow$12.9\%)}}
              & 
              \textbf{12.9\%{\fontsize{6}{10}\selectfont($\downarrow$16.2\%)}} & \textbf{15.7\%{\fontsize{6}{10}\selectfont($\downarrow$5.4\%)}} \\
              \bottomrule
        \end{tabularx}
    \end{table}

    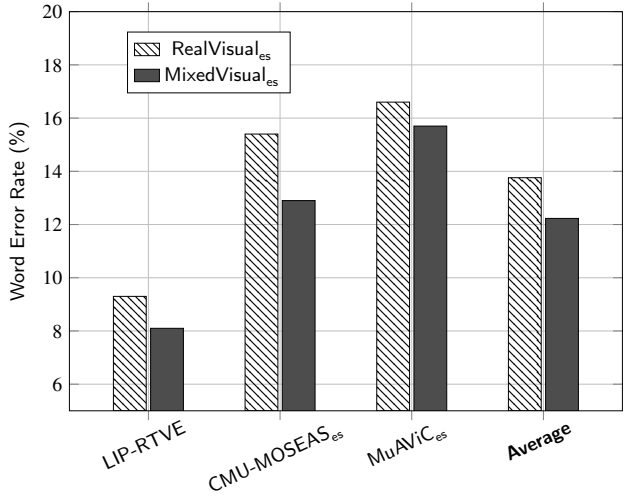
\begin{figure}[pos=t]
      \centering
      \resizebox{\linewidth}{!}{
        \begin{tikzpicture}
          \begin{axis}[
            ybar=2pt, 
            symbolic x coords={LIP-RTVE, CMU-MOSEAS\textsubscript{es}, MuAViC\textsubscript{es}, Average},
            xtick=data,
            xticklabels={LIP-RTVE, CMU-MOSEAS\textsubscript{es}, MuAViC\textsubscript{es}, \textbf{Average}},
            ylabel={Word Error Rate (\%)}, 
            y label style={at={(axis description cs:-0.06,.5)}, anchor=south}, 
            bar width=15pt, 
            enlarge x limits=0.2, 
            ymin=5, ymax=20, 
            grid=major, 
            legend style={at={(0.4,0.95)}}, 
            height=8cm, 
            width=1.25\linewidth, 
            legend image code/.code={%
              \draw[#1] (0.1cm,-0.1cm) rectangle (0.5cm,0.1cm); 
            }, 
            x tick label style={rotate=25, yshift=5pt},
          ]
        
          \addplot[
            fill={rgb, 1:red, 0.8; green, 0.8; blue, 0.8}, 
            pattern=north west lines, 
        ] coordinates {(LIP-RTVE, 9.3) (CMU-MOSEAS\textsubscript{es}, 15.4) (MuAViC\textsubscript{es}, 16.6) (Average, 13.76)};
          
        \addplot[
            fill={rgb, 1:red, 0.3; green, 0.3; blue, 0.3}, 
        ] coordinates {(LIP-RTVE, 8.1) (CMU-MOSEAS\textsubscript{es}, 12.9) (MuAViC\textsubscript{es}, 15.7) (Average, 12.23)};
          
          \legend{RealVisual\textsubscript{es}, MixedVisual\textsubscript{es}}
          
          \end{axis}
        \end{tikzpicture}
      }
      \caption{WER comparison for the audiovisual modality of AVSR models on different datasets. The ``Average'' bar represents the mean WER across the three datasets. The results for the \textit{SynthVisual\textsubscript{es}} model are omitted to focus on the relative improvement achieved through the data augmentation strategy.}
      \label{fig:AVSR_results}
    \end{figure}
    
    As shown in Table~\ref{tab:wer_results}, the model trained on real AV data consistently outperforms the one trained exclusively on synthetic video. However, when synthetic video is used to augment the real AV corpus, the \textit{MixedVisual\textsubscript{es}} model achieves lower WERs across all benchmarks. The gains are particularly notable on LIP-RTVE and CMU-MOSEAS\textsubscript{es}, with relative improvements of 12.9\% and 16.2\%, respectively. The improvement is more moderate on MuAViC\textsubscript{es} (5.4\%), which, as explained in Section~\ref{subsec:lipreading_performance}, is consistent with the higher variability and less controlled recording conditions of this dataset, where visual cues are inherently less informative.
    
    These trends are also illustrated in Figure \ref{fig:AVSR_results}, which shows how augmenting real audiovisual training data with synthetic video consistently improves AVSR performance, although the magnitude of the gain depends on the dataset's visual characteristics.
    
    To contextualize these results, we compare our models directly against the AVSR baselines introduced by \citet{anwar2023muavic}, who first applied the AV-HuBERT framework (the same architecture used in this work) to both multilingual and monolingual training settings. Table~\ref{tab:sota} reports WER results for their multilingual and monolingual Spanish AVSR models alongside our \textit{RealVisual\textsubscript{es}} and \textit{MixedVisual\textsubscript{es}} AVSR models.

    \begin{table}[t]
        \caption{WER comparison of our models against the multilingual and monolingual Spanish AVSR baselines from \citet{anwar2023muavic} on different benchmarks.}
        \label{tab:sota}
        \centering
        \setlength{\tabcolsep}{2.5pt} 
        \fontsize{7}{10}\selectfont
        \begin{tabularx}{\columnwidth}{l c c c}
            \toprule
            \textbf{Model} & \textbf{LIP-RTVE} & \textbf{CMU-MOSEAS\textsubscript{es}} & \textbf{MuAViC\textsubscript{es}} \\
            \midrule
            Anwar et al. (multilingual) & 24.8\% & 25.5\% & 16.2\% \\
            Anwar et al. (monolingual)  & 17.6\% & 16.7\% & 15.9\% \\
            \midrule
            \textit{RealVisual\textsubscript{es}} [Ours] & 9.3\% & 15.4\% & 16.6\% \\
            \textit{MixedVisual\textsubscript{es}} [Ours] & \textbf{8.1\%} & \textbf{12.9\%} & \textbf{15.7}\% \\
            \bottomrule
        \end{tabularx}
    \end{table}

    The comparison first shows that our \textit{RealVisual\textsubscript{es}} model already achieves competitive performance with respect to the AVSR baselines of Anwar et al., outperforming them on two of the three benchmarks while slightly underperforming on MuAViC\textsubscript{es}. This improvement is likely due to the use of a slightly larger real audiovisual training set ($\approx$206h vs. $\approx$180h used by \citet{anwar2023muavic}), suggesting that performance could be further improved with additional real AV data. However, such data is scarce and no larger Spanish AV corpus is currently available, which limits this direction as a viable path for further gains.
    
    To overcome this limitation, we augment the available real audiovisual data with our synthetically generated visual data through the \textit{MixedVisual\textsubscript{es}} training configuration. As a result, \textit{MixedVisual\textsubscript{es}} consistently improves upon \textit{RealVisual\textsubscript{es}} across all three benchmarks, achieving the lowest WER in every case without requiring additional real audiovisual recordings or architectural modifications.
    
\subsection{Comparison of AVSR with Unimodal Baselines}
\label{subsec:unimodal_vs_multimodals}
    To further quantify the contribution of each modality when synthetic visual data is used as a training resource, we conducted a controlled comparison using our best-performing training configuration, \textit{MixedVisual\textsubscript{es}}. Within this configuration, we evaluated the three corresponding model variants, each trained on the same data but using audiovisual (AV), audio-only (A), or video-only (V) input, while keeping all other factors constant.
    
    \begin{table}[t]
        \centering
        \setlength{\tabcolsep}{8pt}
        \caption{WER (\%) comparison between audiovisual (AV), audio-only (A), and video-only (V) models on three Spanish AVSR benchmarks using \textit{MixedVisual\textsubscript{es}}. Relative WER reduction of AV over A is shown in parentheses.}
        \label{tab:unimodal_vs_multimodal}
        \begin{adjustbox}{max width=\columnwidth}
        \begin{tabular}{lccc}
            \toprule
            \textbf{Modality} & \textbf{LIP-RTVE} & \textbf{CMU-MOSEAS\textsubscript{es}} & \textbf{MuAViC\textsubscript{es}} \\
            \midrule
            AV \textsubscript{(AVSR)} & \textbf{8.1\%{\fontsize{6}{10}\selectfont($\downarrow$5.8\%)}} & \textbf{12.9\%{\fontsize{6}{10}\selectfont($\downarrow$4.4\%)}} & \textbf{15.7\%{\fontsize{6}{10}\selectfont($\downarrow$1.3\%)}} \\
            A \textsubscript{(ASR)} & 8.6\% & 13.5\% & 15.9\% \\
            V \textsubscript{(VSR)} & 68\% & 67.1\% & 98\% \\
            \bottomrule
        \end{tabular}
        \end{adjustbox}
    \end{table}
    
    As shown in Table~\ref{tab:unimodal_vs_multimodal}, the audiovisual (AV) model consistently achieves the best performance across all three benchmarks, with relative WER reductions ranging from 1.3\% to 5.8\% compared to the audio-only model. These results suggest that, even under relatively clean acoustic conditions, incorporating visual information can provide modest but consistent improvements in recognition accuracy when synthetic visual data is used during training.
    
    The video-only variant, as expected, performs substantially worse across all benchmarks, with the largest degradation observed on MuAViC\textsubscript{es}, consistent with the challenging visual conditions discussed in Section~\ref{subsec:lipreading_performance}.

\subsection{Input Attribution Analysis in AVSR}
\label{subsec:input_attribution_analysis} 
    To investigate whether training with synthetic visual data influences how the model exploits visual information during inference, we perform an input attribution analysis using Integrated Gradients (IG) \citep{sundararajan2017integratedgradients}. 
    
    Unlike performance metrics such as WER, which only quantify final recognition accuracy, IG provides insight into the model's internal decision process by estimating the contribution of each input feature to its predictions. In the context of AVSR, this allows us to assess whether training with synthetic visual data alters the model's reliance on visual information. If the synthetic visual stream were less informative than real video, the model might learn to place less emphasis on visual cues and rely more heavily on the audio modality.

    IG estimates feature importance by accumulating the gradients of the model output along a continuous interpolation path between a baseline input and the actual sample. Features that consistently produce larger gradients along this path receive higher attribution scores. Following the standard IG formulation, we use zero-valued audio and video inputs as baselines.

    We compute IG attributions for both the audio and video streams of two AV model variants, \textit{RealVisual\textsubscript{es}} and \textit{SynthVisual\textsubscript{es}}, trained with real and synthetic visual data respectively. Although attributions are obtained for both modalities, our analysis focuses on the visual stream, since the objective is to determine whether training with synthetic visual data changes the model's reliance on visual information.
    
    For every utterance in all evaluation benchmarks, IG produces a spatio-temporal attribution map for the video input. To facilitate quantitative comparison across utterances and models, each map is summarized by its mean attribution value, yielding a single video attribution score per sample, where higher values indicate a greater contribution of the visual modality to the model's prediction. For a given model, we denote by $\bar{A}$ the mean video attribution score, averaged across all evaluation samples.

    \begin{figure}[pos=t]
      \centering
      \includegraphics[width=0.95\linewidth]{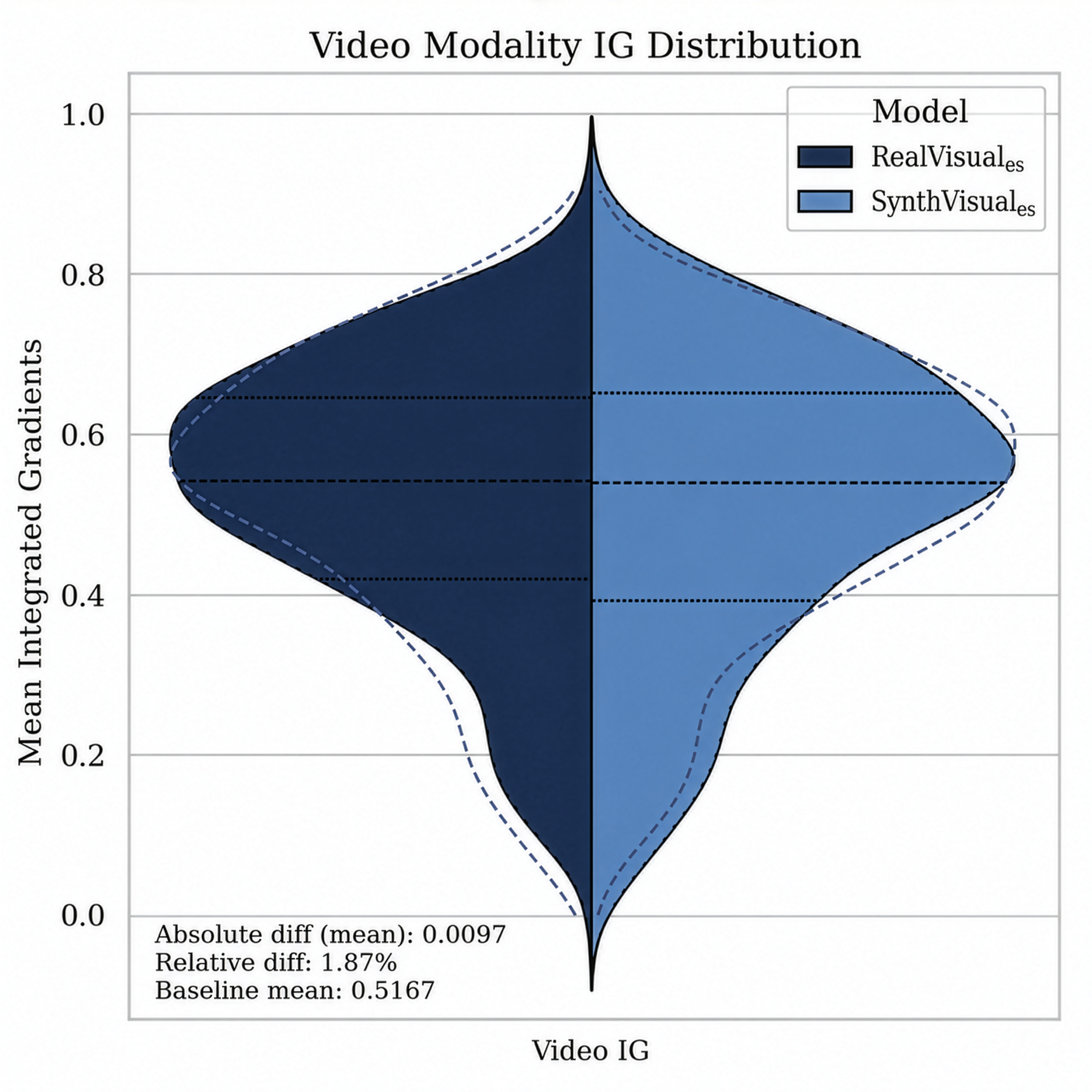}
      \caption{Split violin plot of mean Integrated Gradients for the video modality, comparing \textit{RealVisual\textsubscript{es}} (real video) and \textit{SynthVisual\textsubscript{es}} (synthetic video).}
      \label{fig:ig_violin_video}
    \end{figure}    
    
    Figure~\ref{fig:ig_violin_video} compares the distributions of these sample-level video attribution scores between \textit{RealVisual\textsubscript{es}} and \textit{SynthVisual\textsubscript{es}} using a split violin plot. The distributions largely overlap, with \textit{RealVisual\textsubscript{es}} exhibiting only a marginal 1.87\% higher mean video attribution than \textit{SynthVisual\textsubscript{es}}, computed as $(\bar{A}_{real} - \bar{A}_{synth}) / \bar{A}_{real} \times 100$.

    Both models present nearly identical distribution shapes, as reflected in the close agreement between the dashed outlines representing the mirrored distributions. This negligible difference indicates that models trained with synthetic video retain a comparable level of reliance on visual information, suggesting that synthetic visual data does not bias the model toward predominantly audio-based recognition.

\subsection{Noise Robustness in AVSR} 
\label{subsec:noise_robustness}
    One of the main advantages of audiovisual speech recognition over audio-only systems is its improved robustness under adverse acoustic conditions. To determine whether this benefit is preserved when synthetic visual data is used for training, we evaluated the noise robustness of our best-performing model, \textit{MixedVisual\textsubscript{es}}. Background noise from the MUSAN corpus \citep{musan2015} was added to the audio stream using two noise categories: \textit{``babble''}, simulating background conversations, and \textit{``noise''}, comprising technical and environmental sounds. Each noise type was injected at Signal-to-Noise Ratios (SNRs) ranging from moderately noisy (0 dB) to severely degraded (-20 dB).

    The reported WER values were computed over the combined evaluation data from the three benchmark datasets (LIP-RTVE, CMU-MOSEAS\textsubscript{es}, and MuAViC\textsubscript{es}), providing an overall assessment of the model's robustness.

\begin{figure}[pos=t]
    \centering

    \begin{subfigure}{\columnwidth}
        \centering
        \includegraphics[width=\linewidth]{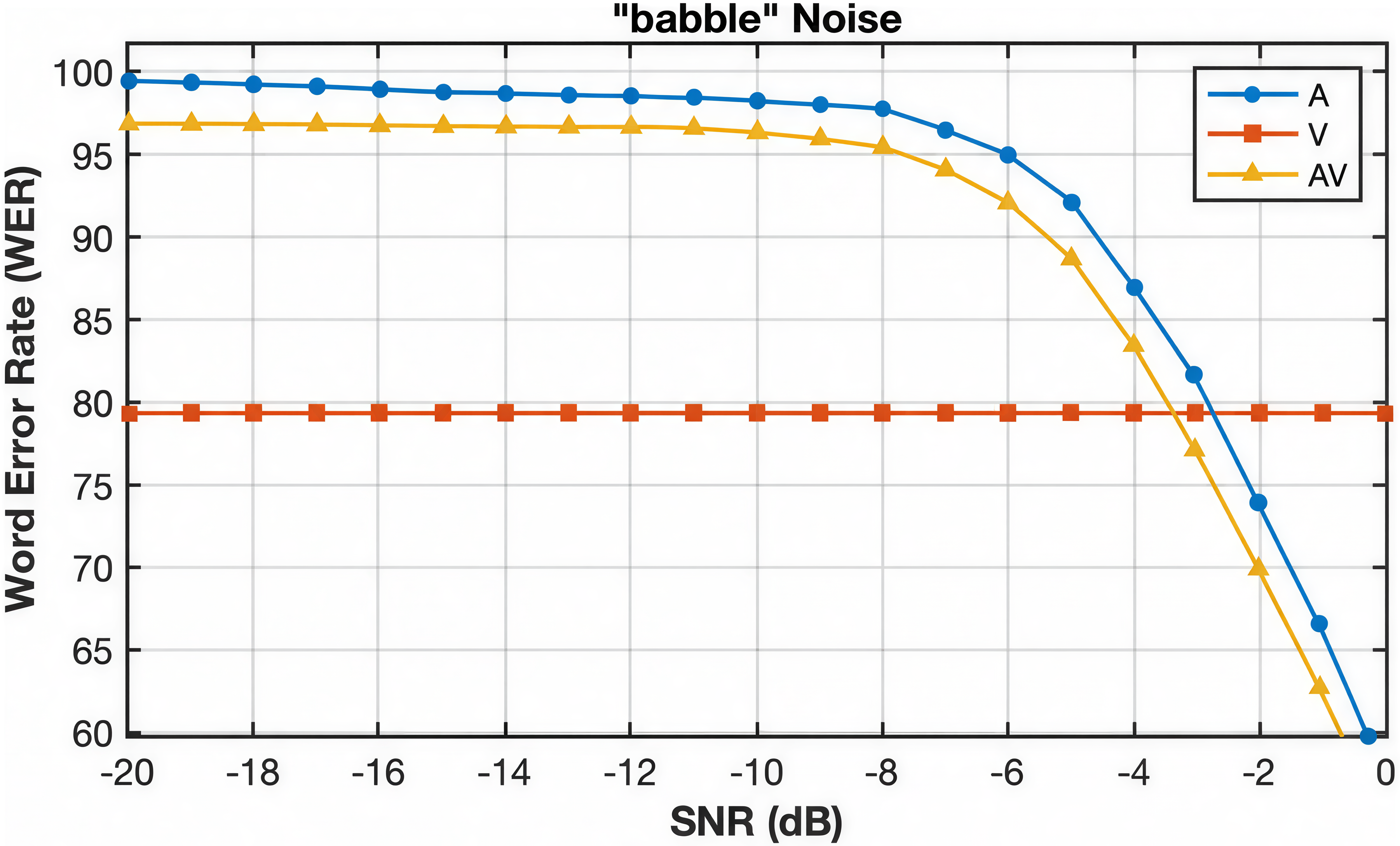}
        \caption{Babble noise.}
        \label{fig:noise_babble}
    \end{subfigure}

    \vspace{0.8em}

    \begin{subfigure}{\columnwidth}
        \centering
        \includegraphics[width=\linewidth]{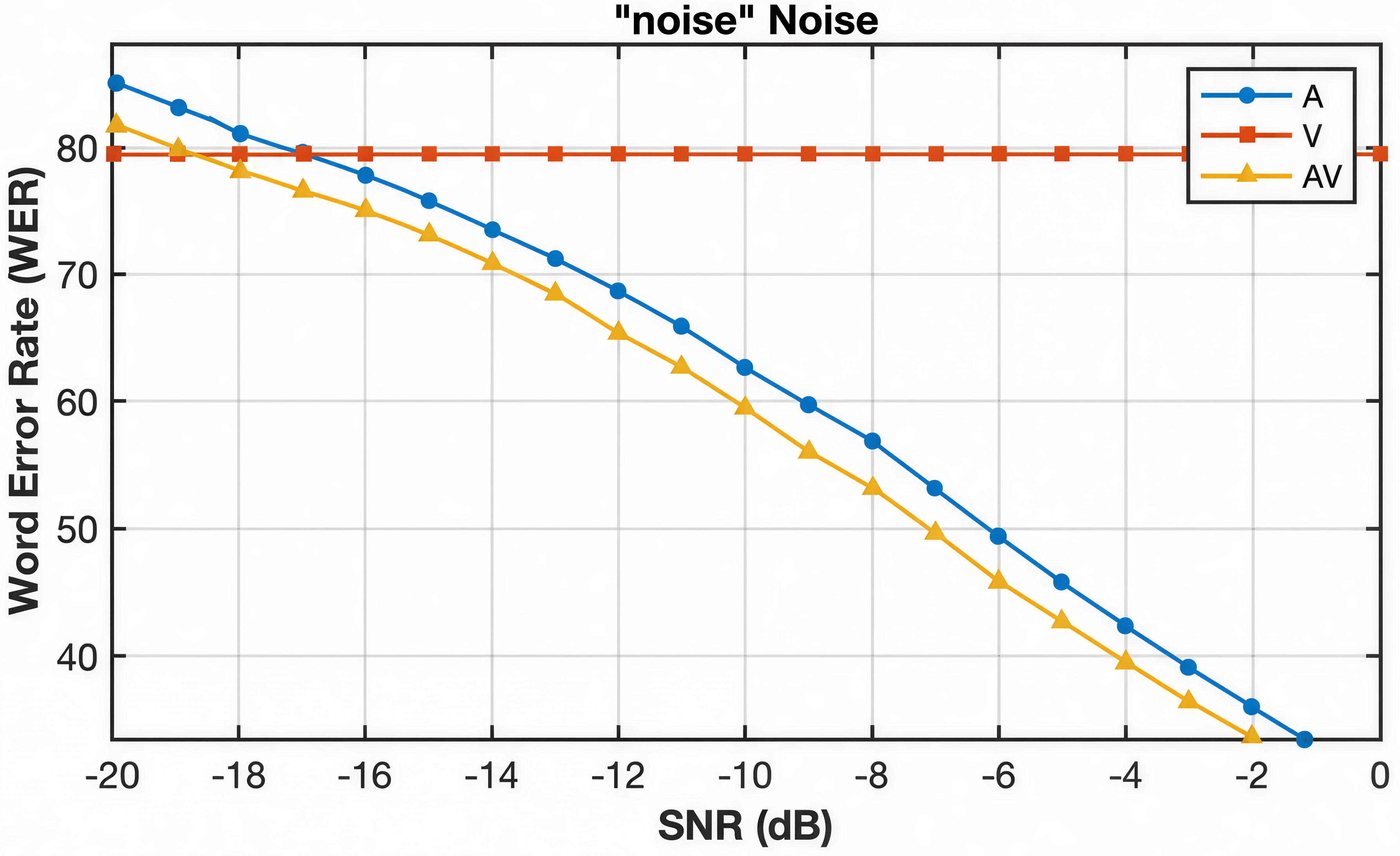}
        \caption{Technical and environmental noise.}
        \label{fig:noise_environmental}
    \end{subfigure}

    \caption{WER performance of \textit{MixedVisual\textsubscript{es}} under different acoustic noise conditions. Results are shown for audiovisual (AV), audio-only (A), and video-only (V) modalities across varying SNR levels.}
    \label{fig:noise_performance_combined}
\end{figure}
    
    The results under babble noise conditions (top panel of Figure~\ref{fig:noise_performance_combined}) show a consistent robustness advantage of the audiovisual (AV) model over the audio-only (A) model. Between 0 dB and -20 dB SNR, the AV model achieves relative WER improvements ranging from approximately 5\% to 10\%, with the gap generally increasing as the acoustic conditions become more challenging. Only at very low SNRs (below -3 dB), where babble noise severely degrades the audio signal due to its spectral similarity to speech, does the video-only (V) model slightly outperform AV, indicating that the audio modality contributes little useful information under these conditions.

    Under technical and environmental noise (bottom panel of Figure~\ref{fig:noise_performance_combined}), a similar overall trend is observed. The AV model consistently outperforms the audio-only model, although the relative improvements are slightly smaller than under babble noise, remaining above 4\% in most conditions. Unlike babble noise, technical and environmental noise degrades the audio modality less severely, allowing both the AV and A models to maintain better performance over a wider range of SNRs. Consequently, the video-only model does not outperform the AV model until approximately -18 dB, substantially later than in the babble noise setting.
    
    Overall, these results indicate that models trained with synthetic visual data preserve the characteristic robustness of audiovisual speech recognition under adverse acoustic conditions. Although the gains over audio-only recognition are more modest than those reported for comparable systems trained on real audiovisual data~\citep{anwar2023muavic}, they remain consistent across both noise types and a wide range of SNRs, indicating that training with synthetic visual data enables the model to exploit complementary visual information when the acoustic signal is degraded.
    
\subsection{Zero-AV-Resource Adaptability}
\label{subsec:zero_resource_language_adaptability} 
    After demonstrating the effectiveness of synthetic visual data as an augmentation strategy for AVSR in Spanish, a language with limited but existing audiovisual resources, we extend this methodology to a more challenging scenario in which no real labeled AV data are available for training. Catalan serves as a representative case study, as no publicly available labeled AV corpora currently exist for the language.
    
    To this end, we defined the \textit{SynthVisual\textsubscript{cat}} training configuration (Table~\ref{tab:trained_models}) using only synthetic AV data generated with the pipeline presented in Section~\ref{subsec:pipeline}. Since the objective is to evaluate generalization to real AV speech, the models were evaluated on our manually annotated benchmark of real Catalan AV recordings, AV-CAT, created specifically for this work due to the absence of an existing evaluation dataset. The resulting performance for the three model variants is summarized in Table~\ref{tab:results_cat}.
    
    \begin{table}[t]
        \centering
        \caption{WER of the \textit{SynthVisual\textsubscript{cat}} model, evaluated on the manually labeled AV-CAT benchmark. Results are shown for audiovisual (AV), audio-only (A), and video-only (V) trained variants.}
        \label{tab:results_cat}
        \begin{tabular}{@{}lccc@{}}
            \toprule
            \textbf{Model / Modality} & \textbf{Audiovisual} & \textbf{Audio-only} & \textbf{Video-only} \\
            \midrule
            \textit{SynthVisual\textsubscript{cat}} & \textbf{19.6\%} & 23.1\% & 105\% \\
            \bottomrule
        \end{tabular}
    \end{table}

    The results show that the audiovisual (AV) model trained entirely on synthetic data outperforms both unimodal baselines, confirming the benefit of multimodal integration even in a zero-AV-resource context. Specifically, it achieves a relative WER reduction of 15.2\% compared to the audio-only model, and 81.3\% compared to the video-only model.
    
    Although the overall WER is slightly higher than that achieved by the Spanish \textit{MixedVisual\textsubscript{es}} model (Table~\ref{tab:unimodal_vs_multimodal}), the multimodal advantage is consistently preserved. This performance gap is expected, as the Catalan model relies exclusively on synthetic audiovisual training data and therefore cannot benefit from the diversity and quality of real audiovisual recordings.
    
    Despite being the first AVSR model for Catalan, \textit{SynthVisual}\allowbreak\textsubscript{cat} achieves promising performance, supporting the feasibility of developing AVSR systems based solely on synthetic video and demonstrating the potential of this approach in low-resource scenarios where audiovisual corpora are not available.

\section{Conclusions}
    This work investigated the use of synthetic visual data to address the scarcity of labeled audiovisual resources for audiovisual speech recognition (AVSR). By generating lip-synchronized visual speech from existing audio recordings, we explored two complementary scenarios: augmenting limited real audiovisual datasets and enabling AVSR training when no real labeled audiovisual data are available.
    
    Our experimental results demonstrate that synthetic visual data constitute an effective augmentation strategy, consistently improving AVSR performance across all Spanish benchmarks, with relative WER reductions of up to 16.2\%, while maintaining the robustness advantages of multimodal speech recognition under noisy conditions. Furthermore, our Catalan case study suggests that synthetic visual data alone can provide sufficient cross-modal information to train an AVSR system in a zero-AV-resource setting, highlighting the potential of this approach to extend AVSR to languages lacking labeled audiovisual corpora.
    
    Overall, the proposed methodology reduces the dependence on language-specific audiovisual datasets and provides a scalable strategy for training AVSR systems in low-resource scenarios. By leveraging existing speech corpora to generate synthetic visual data, it facilitates the development of multimodal speech recognition systems and broadens their accessibility across languages.

\section*{Funding}
This research did not receive any specific grant from funding agencies in the public, commercial, or not-for-profit sectors.

\printcredits

\section*{Declaration of competing interest}
The authors declare that they have no known competing financial interests or personal relationships that could have appeared to influence the work reported in this paper.

\section*{Acknowledgements}
Not applicable.

\section*{Data availability}
The publicly available datasets used in this study can be obtained from the original sources cited throughout the manuscript. The implementation of the proposed methodology, together with the scripts required to reproduce the experiments, is publicly available at \url{https://github.com/Pol-Buitrago/SynthAVSR}.

The synthetic audiovisual datasets and the manually annotated Catalan audiovisual benchmark generated during this study are not publicly available because they are derived from third-party audiovisual content that cannot be redistributed. Additional information may be provided by the corresponding author upon request, subject to the licensing restrictions of the original data sources.

\bibliographystyle{elsarticle-harv}
\bibliography{custom}

\end{document}